\documentclass[preprint,12pt]{aastex}

\def\eg{{e.g.,}}

\def\etal{{et~al.\null}}
\def\ie{{i.e.,}}

\def\kms{{km~s$^{-1}$}}
\title{A Survey for Planetary Nebulae in M31 Globular Clusters}
\begin{document}
\shorttitle{Planetary Nebulae in M31 Globular Clusters}
\author{George H. Jacoby}
\affil{Giant Magellan Telescope / Carnegie Observatories, 
813 Santa Barbara Street, Pasadena, CA 91101}
\email{gjacoby@gmto.org}
\author{Robin Ciardullo}
\affil{Department of Astronomy \& Astrophysics, The Pennsylvania State
University,  University Park, PA 16802}
\email{rbc@astro.psu.edu}
\author{Orsola De Marco}
\affil{Department of Physics \& Astronomy, Macquarie University, Sydney, 
NSW 2109, Australia}
\email{orsola.demarco@mq.edu.au}
\author{Myung Gyoon Lee}
\affil{Astronomy Program, Department of Physics and Astronomy, Seoul 
National University, Seoul 151-742, Korea}
\email{mglee@astrog.snu.ac.kr}
\author{Kimberly A. Herrmann}
\affil{Lowell Observatory, Flagstaff, AZ 86001}
\email{herrmann@lowell.edu}
\author{Ho Seong Hwang}
\affil{Smithsonian Astrophysical Observatory, 60 Garden Street, Cambridge,  
MA 02138}
\email{hhwang@cfa.harvard.edu}
\author{Evan Kaplan}
\affil{Vassar College, 124 Raymond Ave., Poughkeepsie, NY 12604}
\email{evanskaplan@gmail.com}
\and
\author{James E. Davies}
\affil{Smithsonian Astrophysical Observatory, 60 Garden Street, Cambridge,
MA 02138}
\email{jdavies@cfa.harvard.edu}

\begin{abstract}
We report the results of an [O~III] $\lambda 5007$ spectroscopic survey 
for planetary nebulae (PNe) located within the star clusters of M31.
By examining $R \sim 5000$ spectra taken with the WIYN+Hydra spectrograph, 
we identify 3 PN candidates in a sample of 274 likely globular clusters, 
2 candidates in objects which may be globular clusters, and 5 candidates
in a set of 85 younger systems.   The possible PNe
are all faint, between $\sim 2.5$ and 
$\sim 6.8$~mag down the PN luminosity function, and, partly as a 
consequence of our selection criteria, have high excitation, with [O~III] 
$\lambda 5007$ to H$\beta$ ratios ranging from 2 to $\gtrsim 12$.   We 
discuss the individual candidates, their likelihood of cluster membership,
and the possibility that they were formed via binary interactions 
within the clusters.  Our data are consistent with the suggestion 
that PN formation within
globular clusters correlates with binary encounter frequency, though, due to
the small numbers and large uncertainties in the candidate list, this
study does not provide sufficient evidence to confirm the hypothesis.

\end{abstract}

\section{Introduction}
Despite over a century of observations and decades of detailed
modeling, the stellar population that forms planetary nebulae (PNe) is still 
somewhat of a mystery.  The traditional theory states that the progenitors 
of PNe are low- to intermediate-mass single stars at the end of the AGB phase
\citep{shklovski, abell}.  This hypothesis explains the distribution of
PNe throughout space, and is responsible for the widely-held belief that
the Sun will eventually evolve through this easily identifiable nebular phase
\citep[\eg][]{abell,p2,buzzoni}.  However, this theory does not provide
a natural explanation for the non-spherical morphologies observed for the
great majority of PNe, nor their low rate of formation.  For these and other
inconsistencies \citep[see][]{demarco09}, a new paradigm has been developed,
wherein most planetary nebulae are shaped via the interaction of an AGB wind 
with a binary companion \citep[\eg][]{soker97}.

Unfortunately, while the binary interaction model explains some of the 
anomalies associated with the observed planetary nebula population
\citep{moe, demarco09, moe12}, this theory awaits final confirmation:  the number of
PN central stars with known binary companions is still relatively 
small, and programs to detect such objects are extremely challenging
\citep{demarco11, demarco12}.  Moreover, neither the single-star 
nor binary-star hypothesis can explain the luminosities observed for PNe in
the old stellar populations of elliptical galaxies and spiral bulges
\citep{straggler}, nor the invariance of the bright end of the planetary
nebula luminosity function (PNLF) across stellar populations \citep{p11}.  
For this, one must invoke yet another formation scenario, wherein 
significant mass transfer (or a complete stellar merger) occurs
prior to the planetary nebula phase \citep{straggler, symbiotic}.

It is difficult to probe the different PN formation scenarios using field
stars, since one has almost no prior information about the properties of the
PN progenitors.  However, within star clusters, the situation is different,
as both the age and metallicity of the progenitor can be accessed.  
Moreover, the low turnoff mass of old globular clusters (GCs)
provides a tool with which to probe the binary formation scenario 
directly.  Because GCs generally have turnoff masses less than $1 M_{\odot}$, 
any post-AGB core arising from simple single-star stellar evolution must have 
a very small mass, $M_{\rm core} \lesssim 0.54 M_{\odot}$ \citep{kalirai}, and 
an extremely long evolutionary timescale, $t > 10^5$~yr \citep[\eg][]{caloi}.  
Such objects cannot make planetary nebulae by themselves, since the mass lost 
during the AGB phase will disperse long before the core becomes hot
enough to generate ionizing radiation.   Any PN detected in these systems must 
therefore come from an alternate evolutionary channel, such as 
a common-envelope interaction or a mass augmentation process (i.e., a 
stellar merger).

Searches for planetary nebulae in Galactic clusters have turned up only
a few associated objects.  Although more than a dozen PNe are projected
near open clusters, the vast majority are undoubtedly line-of-sight 
coincidences \citep{majaess, parker}.  Similarly, out of 130~Galactic GCs 
surveyed, only four host PNe:  Ps~1 in M15 \citep{pease}, GJJC-1 in M22 
\citep{gillett}, JaFu1 in Pal~6, and JaFu2 in NGC~6441 \citep{jacoby+97}.  Two 
of these PNe have high mass central stars more appropriate to PNe 
within open clusters \citep[$\sim 0.62 \, M_{\odot}$ for Ps~1 and 
$\sim 0.75 \, M_{\odot}$ for GJJC-1;][]{bianchi, harrington}, while the others 
have highly non-spherical nebulae \citep{demarco_apn}.  (The true 
nature of GJJC-1 is currently being re-assessed due to its high stellar
mass, low nebular mass, and bizarre chemical composition \citep{jacoby12},
but for this paper, we adopt the usual PN classification.)  
These facts, along with 
the observation that three of the four PNe are located in clusters that are 
rich in X-ray sources, suggest that interacting binaries play a role in the 
formation of cluster PNe \citep{jacoby+97}.

Since the sample of PNe within Galactic globular clusters is small, the 
significance of any conclusion based on their properties is low.  To better 
understand the processes that form PNe within clusters, many more objects are 
required.  For this, we must look to other galaxies.  Unfortunately, while
there have been a few isolated associations of [O~III] $\lambda 5007$ emission
with extragalactic globular clusters \citep{minniti, bergond, larsen, 
chomiuk}, these discoveries have largely been serendipitous, and
in most cases only sensitive to the most luminous planetary nebulae.
\citet{peacock-4472} did conduct a systematic search for PNe within the 
massive globular clusters of the giant Virgo elliptical NGC~4472, but again, 
this survey (which did not find any objects) was only sensitive to objects in 
the top $\sim 2.5$~mag of the luminosity function.   Since bright PNe are 
relatively rare, a deeper investigation is needed to place better
constraints on the phenomenon.

Here we report the results of a spectroscopic survey for planetary nebulae 
within the star clusters of M31.  In \S 2, we describe our multi-fiber
observations and the basic reduction steps needed to analyze
$\sim 460$ candidate star clusters in M31.  In \S 3, we describe our search
for embedded planetary nebulae, and the techniques required to recover
objects that are more than $\sim 6$~mag down the planetary nebula luminosity 
function.  We present a list of the clusters surveyed, along with their
magnitudes and radial velocities, and identify those clusters in which [O~III] 
$\lambda 5007$ emission is present.  In \S 4, we describe the individual PN 
candidates and their host clusters.  Finally, we discuss our results, 
and show that the number of PNe recovered within M31's globular cluster
system is roughly consistent with surveys of Milky Way clusters.



\section{Observations and Reductions}
On 2008 Oct 25-28 (UT), we targeted 467 candidate star clusters in M31 with 
the 3.5-m WIYN telescope on Kitt Peak and the Hydra multi-fiber spectrograph. 
The objects selected for study were largely taken from a list of clusters 
given in the Revised Bologna Catalog \citep[RBC;][]{galleti04, galleti07}
and supplemented using the X-ray cluster identifications of
\citet{fan} and the young cluster candidates of \citet{caldwell09}.
As a control, we also positioned spare fibers on known M31
planetary nebulae taken from the list of \citet{merrett}.  These PNe
span a range of brightnesses, from $20.5 < m_{5007} < 25.9$ 
(\ie\ $-13.71 > \log F_{5007} > -15.85$), and allowed us to test
the depth of our exposures and the accuracy of our wavelength calibration.
Finally, to estimate the background light and the amount of diffuse 
[O~III] emission arising from M31's disk, several fibers in each setup were 
offset onto regions of blank sky.

\begin{figure}[t]
\figurenum{1}
\plotone{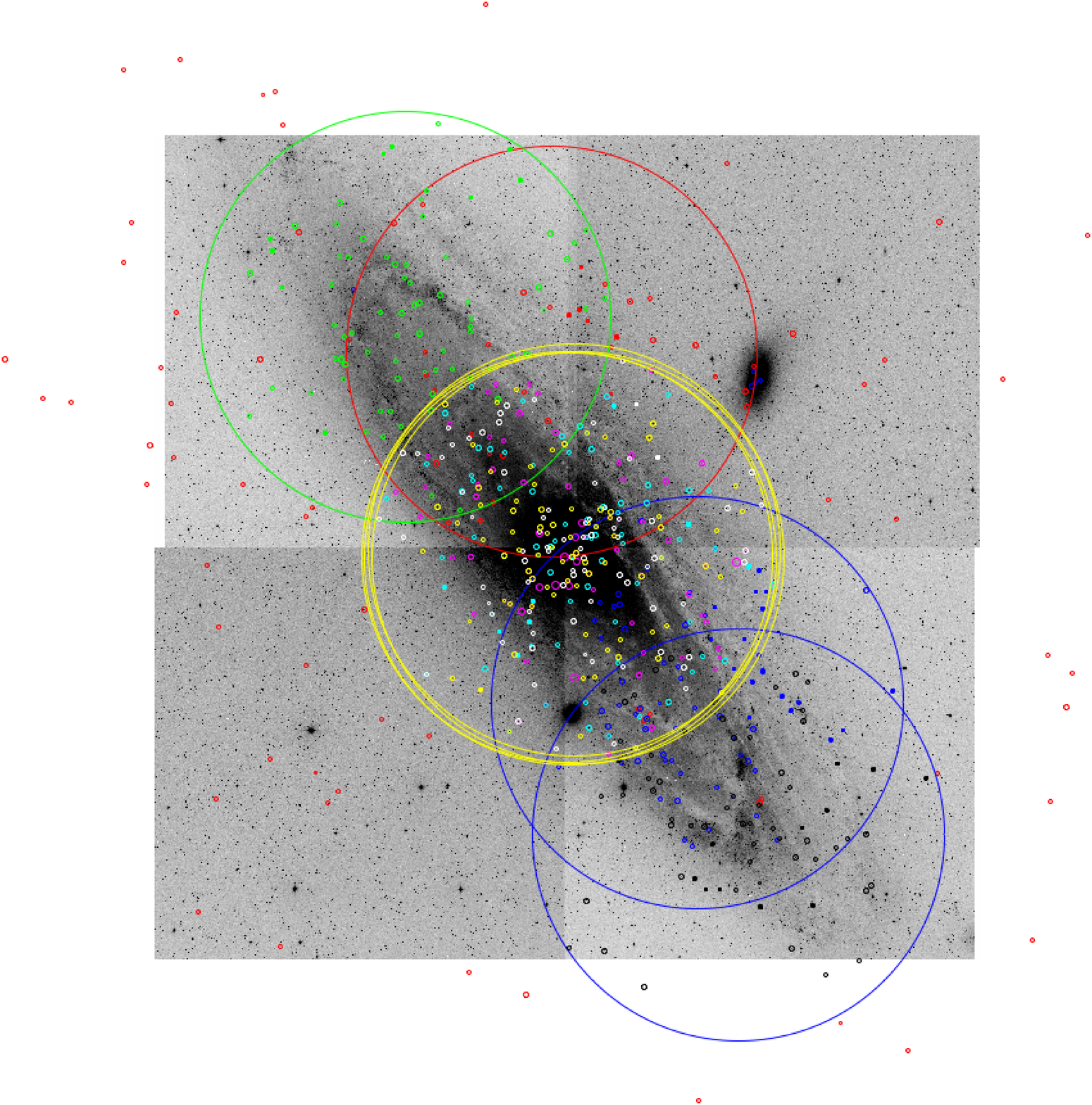}
\caption[M31_fields]{The locations of our WIYN+Hydra fields and the targeted
M31 clusters, superposed on a mosaic of [O~III] images from \citet{massey}.  
North is up, and east is to the left.  Each $1^\circ$ colored circle represents 
a different Hydra setup.}
\label{M31_fields}
\end{figure}

Our specific target selections were made using the fiber-assignment
program {\tt whydra}.  Since the M31 globular cluster system covers a
much larger area than the $1^{\circ}$ field-of-view of the WIYN+Hydra 
instrument (see Figure~\ref{M31_fields}), we began by visualizing the locations
of the individual clusters using the [O~III] images of the Local Group Galaxies 
Survey \citep{massey}.  To accommodate the bulk of the galaxy's cluster
population, we located our first four setups on M31's bulge.
Thereafter, we alternated between regions southwest and
northeast of the galactic center, each time allowing {\tt whydra} to
identify an optimal field position by searching a $25 \times 25$ grid of
space in $0\farcm 1$ increments, with 15 possible rotation angles (between 
$0^\circ$ and $30^\circ$) at each location.  Top priority for fiber assignments 
was always given to previously unobserved clusters, with duplicate clusters,
field PNe, and blank sky positions assigned lesser precedence.  In total,
eight different setups were executed over the four nights of the observing run,
and data were acquired for 391 RBC clusters, 64 X-ray clusters, and 12 young
clusters, with 30 of the systems being targeted more than once.  An 
additional 55 of M31's field planetary nebulae were also observed.
A log of the observations is given in Table~\ref{obslog}.  

To execute these observations, the WIYN+Hydra system was configured to use
the instrument's array of $3\arcsec$ diameter blue-sensitive fibers, the WIYN 
Bench spectrograph, and a 740~lines~mm$^{-1}$ Volume Phase Holographic (VPH) 
grating designed to optimize throughput near 5000~\AA.  The resultant
spectra covered the wavelength range between 4400~\AA\ and 5450~\AA\ at
1~\AA\ resolution and 0.5~\AA~pixel$^{-1}$.  Typically, each Hydra setup was 
observed for 3.5~hr via a series of seven 30 minute exposures.  

Our initial reduction procedures were similar to those described
in \citet{kim2}.  We began with the tasks within the 
IRAF\footnote{IRAF is distributed by the National Optical Astronomy 
Observatory, which is operated by the Association of Universities for 
Research in Astronomy (AURA) under cooperative agreement with the 
National Science Foundation.} {\tt ccdred} package: 
the data were trimmed and bias-subtracted via {\tt ccdproc}, dome flats 
(typically three per setup) were combined using {\tt flatcombine}, and 
CuAr comparison arcs, which bracketed the program exposures, were combined via 
{\tt imcombine}. Next, {\tt dohydra} within the {\tt hydra} package was used 
to combine and linearize the spectra, with the averaged dome flats serving 
to define the extraction apertures, and the averaged comparison arcs 
providing the wavelength calibration.  We estimate these wavelength 
calibrations to be precise to 0.04~\AA\ yielding $1\,\sigma$ errors of 
$\sim 2.4$~\kms.  As we illustrate below, this is small in comparison
with the other uncertainties associated with our radial velocity 
determinations.

After extracting each spectrum, the program objects were sky subtracted 
using the data acquired through our blank-field fibers.  This step was
straightforward: since all the data were taken in dark time and no bright
sky lines fell within the wavelength range of the instrument, we simply
used {\tt scombine} to combine the extracted spectra from the multiple
exposures and then invoked {\tt skysub} to perform the subtraction.  As will
be explained in \S~3, our searches for PN emission involved the subtraction
of template spectra derived from the data themselves.  Since these 
templates underwent the same sky subtraction, our procedure was 
differential in nature, and the details of sky subtraction had no
significant effect on the analysis.

The final step in the basic data reduction involved estimating the response
function of our instrument.  Relative spectrophotometry across the entire
spectral range of WIYN+Hydra was unnecessary for our program.  
Instead, we concentrated on determining the instrumental sensitivity at 
H$\beta$ and [O~III] $\lambda 4959$ relative to [O~III] $\lambda 5007$\null.
The [O~III] $\lambda 4959$ calibration was straightforward.  Of the
55 planetary nebulae targeted via our spare fibers, all but one
were recovered in [O~III] $\lambda 5007$, and 51 out of 55 had
detectable [O~III] $\lambda 4959$.  By comparing the observed ratio
of the oxygen doublet to the astrophysical ratio of 2.92, we concluded 
that the spectroscopic throughput at 4959~\AA\ was $\sim 90\%$ that at
5007~\AA.

The procedure for estimating the instrumental response at H$\beta$ was 
slightly more complicated.  We began by identifying 15 bright clusters and 
comparing their continuum flux near 5000~\AA\ to that surrounding the H$\beta$ 
line.  Spectral libraries \citep[\eg][]{jacoby+84} show that, after applying
M31's foreground reddening \citep[$E(B-V) = 0.062$;][]{schlegel}, the
spectral energy distributions of old stellar populations should be 
flat between 4800~\AA\ and 5100~\AA\null.  Consequently, to estimate the 
relative system response at H$\beta$, we simply adopted the inverse of the
observed $\lambda 5007$ to H$\beta$ continuum ratio.  Our estimate of 
68\% is slightly greater than the 62\% value expected from the Bench
spectrograph system and the VPH grating \citep{bershady}.  However, since the 
efficiency of this grating is only known to $\sim 10\%$, the two numbers
are consistent.

\begin{figure}
\figurenum{2}
\plotone{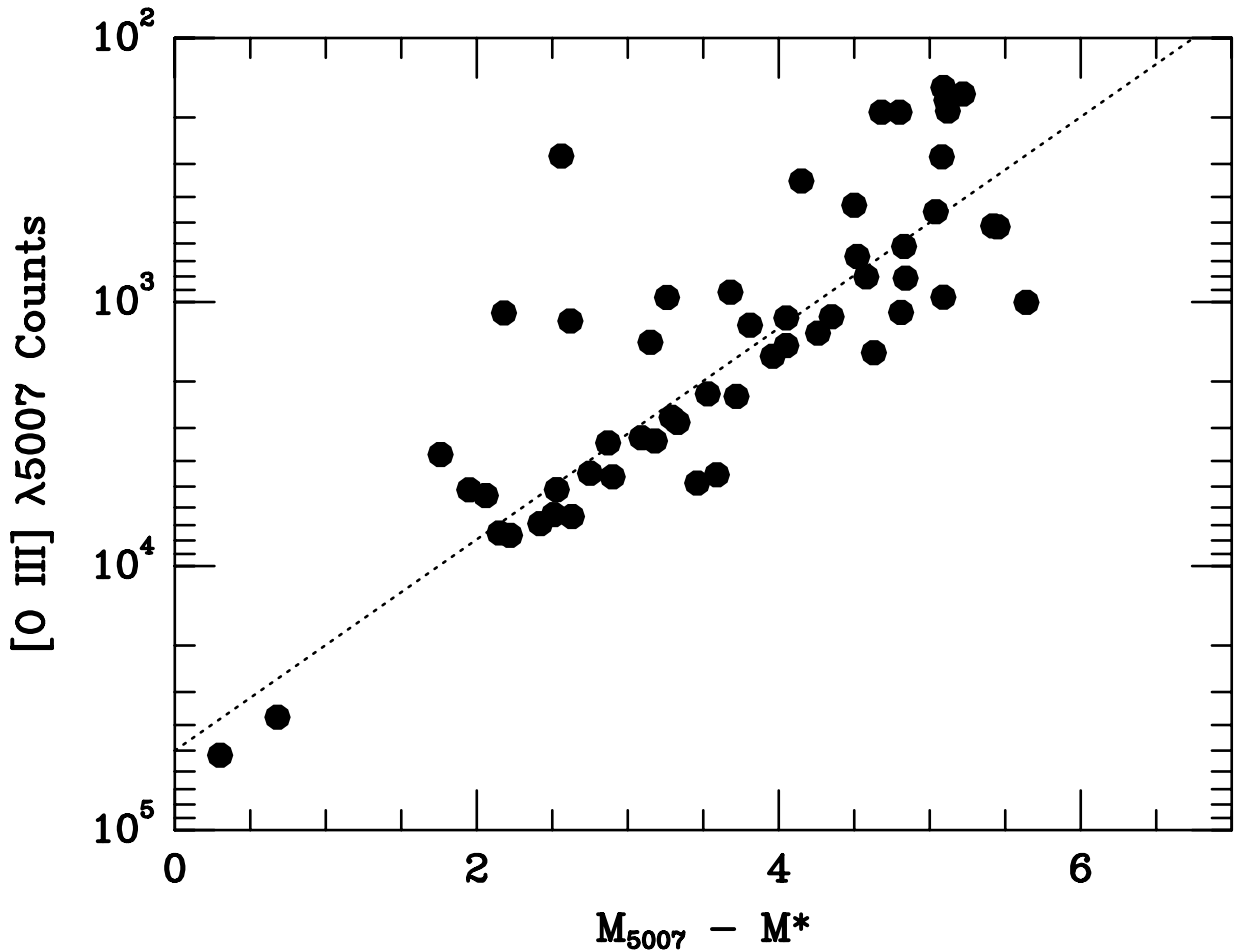}
\caption[pn_counts]{A comparison of the absolute [O~III] magnitudes of M31 PNe 
measured by \citet{merrett} to the number of $\lambda 5007$ counts recorded in 
our spectra.  The scatter is due to imperfect astrometry and fiber positioning
errors, and the effect of clouds on one of the Hydra setups.  The dotted
line shows the one-to-one relation.  The figure illustrates that we should
be able to detect PNe that are more than $\sim 6$~mag down the luminosity
function.  For reference, $M^* = -4.5$, which in M31 is equivalent
to a monochromatic flux of $2.9 \times 10^{-14}$~ergs~cm$^{-2}$~s$^{-1}$.}
\label{pn_counts}
\end{figure}

Figure~\ref{pn_counts} compares the [O~III] $\lambda 5007$ counts recorded 
in our PN spectra to the objects' apparent magnitudes as determined by the
counter-dispersed imaging of \citet{merrett}.  There is a substantial amount 
of scatter in the diagram, due to the photometric and astrometric errors 
associated with the measurements from the Planetary Nebula Spectrograph, 
imperfections in our fiber positioning, and the effects of variable sky 
conditions.  Nevertheless, the data demonstrate that our observations go quite 
deep: since we can detect emission lines containing as few as 
$\sim 100$~counts, our observations are sensitive to PNe that are
$\sim 6$~mag down the luminosity function.

\section{Finding PNe Within Star Clusters}
Three criteria must be met before we can claim the detection of a PN candidate
within an M31 star cluster.  First, we must identify the presence of emission
lines in the spectrum of the cluster.  Second, the observed emission lines 
must have line-ratios consistent with those expected from a planetary
nebula.  Because of our limited spectral coverage (chosen to achieve the
resolution needed to optimize
emission line detections), this criterion is equivalent to requiring 
that the ratio of [O~III] $\lambda 5007$ to H$\beta$ be greater than some 
threshold (see below).  Third, the velocity of a PN candidate, as derived from 
its emission-lines, must be consistent with that of a star bound to its parent
star cluster.  Since the escape velocity from a typical M31 cluster is 
low, the precision of our emission-line and absorption line velocity 
measurements is an important parameter for our selection criteria.

\subsection{Detection of Emission-lines}
\begin{figure}[t]
\figurenum{3}
\epsscale{0.6}
\plotone{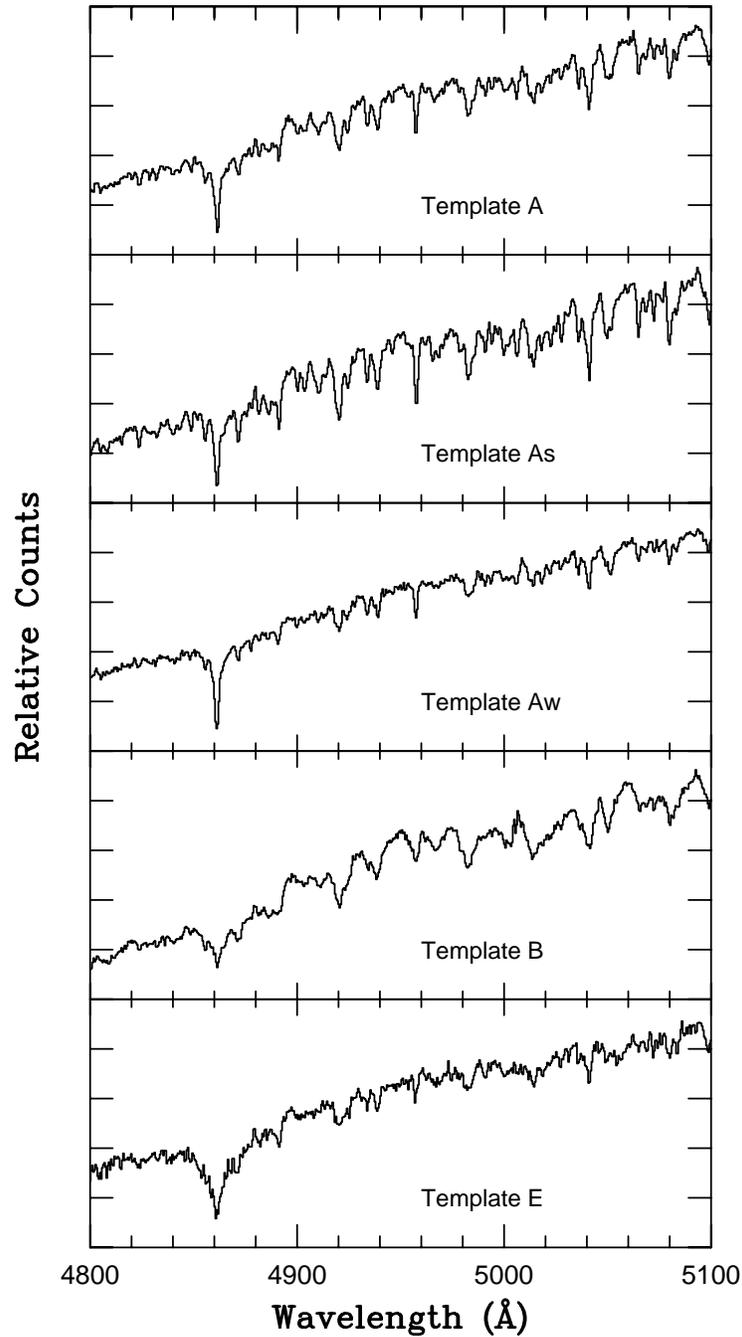}
\caption[templates]{Spectra of the five template clusters used to suppress the 
stellar continuum and enhance the visibility of emission lines.   The strongest
absorption in this part of the spectrum comes from H$\beta$; most of the 
others lines are due to iron.   The differences between spectra represent 
variations in cluster's age and metallicity.}
\label{templates}
\end{figure}

To search for PN emission within globular clusters, we began by
normalizing each program spectrum with the {\tt continuum}
command within IRAF\null.   We next divided the spectra into five
classes based on the strengths and widths of the absorption features,
thereby effectively grouping the clusters by age and metallicity.
The highest signal-to-noise spectrum of each class was then chosen as a
template, and shifted to zero velocity using, as a reference, the absorption
lines of H$\beta$, the Mgb triplet ($\lambda\lambda$5167,5173,5184), 
and other strong features.  The remaining
clusters of the class were then cross-correlated against their
template to determine their relative radial velocities, and shifted to the
rest frame.  Finally, to improve the detectability of any faint
emission feature which may be lost amidst the background light, the 
template spectra were scaled and subtracted from the other members of 
their class.  Figure~\ref{templates} shows the five template spectra.

\begin{figure}
\figurenum{4}
\epsscale{0.6}
\plotone{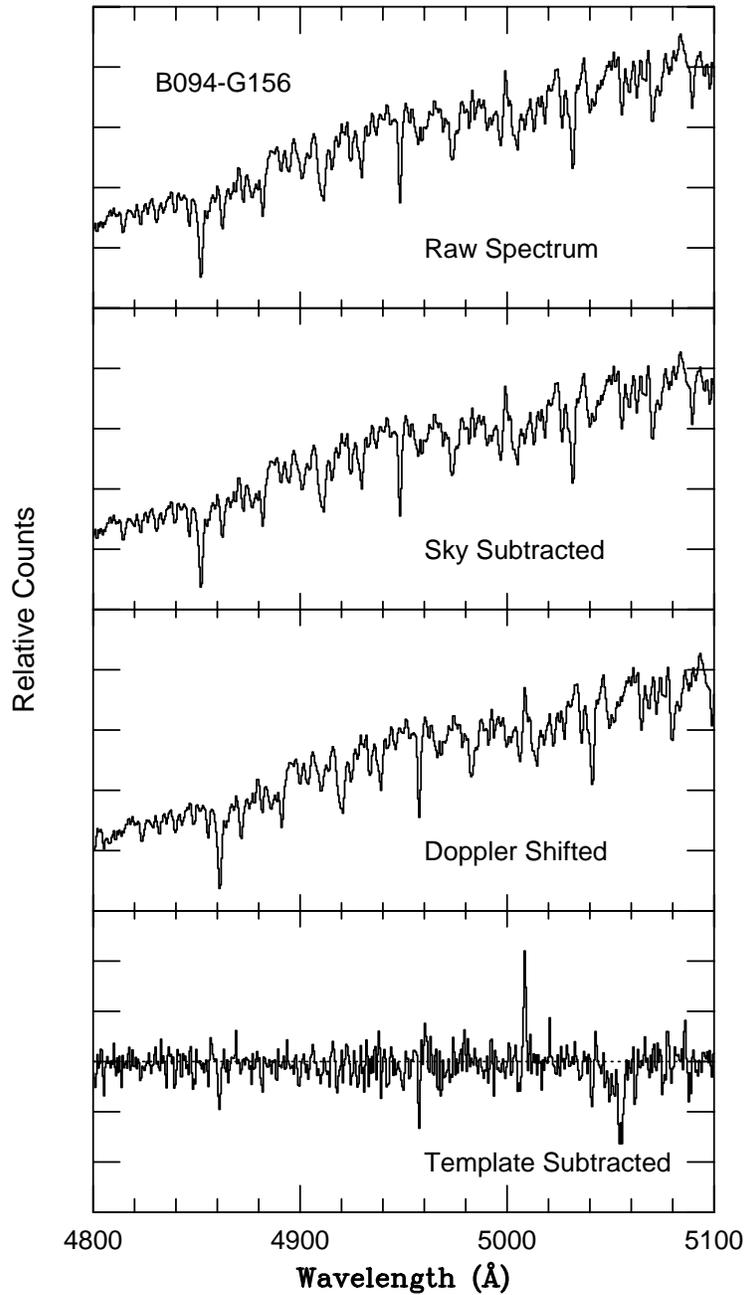}
\caption[example_spec]{The spectra of an M31 globular cluster between 
4800~\AA\ and 5100~\AA, showing our raw data, along with the data after sky
subtraction, Doppler shifting, and template subtraction.  Note that after 
template subtraction, the [O~III] $\lambda 5007$ emission line is easily
seen.  This is a well-detected and well-measured emission-line source 
containing $\sim 1000$~counts in [O~III] $\lambda 5007$.}
\label{example_spec}
\end{figure}

Figure~\ref{example_spec} illustrates this template-subtracting procedure
using the globular cluster B094-G156\null. This example is representative 
of our reductions; some of the template subtractions are much better, while 
others are poorer (see Section 4).  In particular, because there were only five
template clusters, not all the observed systems flattened as well as the 
one that is displayed.  In particular, several of the clusters classified 
as ``young'' by \citet{caldwell09} had imperfect subtractions around H$\beta$.
Nevertheless, in virtually all cases, the technique worked well around
5007~\AA, as it effectively suppressed the continuum, allowing us to
detect extremely weak emission from [O~III].  Moreover, even when 
H$\beta$ was poorly subtracted, we could still measure the emission-line 
ratios extremely well, as the underlying stellar absorption was far broader 
than the unresolved Balmer emission.

\subsection{The Emission-line Signature}
Our next step was to look for evidence of planetary nebula emission. 
Because our data were taken using a multi-fiber 
spectrograph, rather than traditional slit spectroscopy, local sky subtraction 
was not possible.  Consequently, the emission arising from the diffuse
ionized gas of M31's disk could not always be removed cleanly, and, even
in the galactic bulge, line contamination was frequently a problem
\citep[see][for an image of M31's bulge emission]{ciardullo88}.  In fact, 
as summarized in Table~\ref{gc_summary}, roughly one-third of the globular 
clusters surveyed displayed some evidence of emission, due mostly to M31's 
warm interstellar medium.

To guard against this form of false detection, we considered the 
expected line ratios of a planetary nebula.  Most bright PNe are high 
excitation objects: in the top 2.5 mag of the [O~III] PNLF, all planetary 
nebulae have [O~III] $\lambda 5007$ to H$\alpha$ ratios greater than 3
(\ie\ $I(\lambda 5007)/I({\rm H}\beta) \gtrsim 9$), and even at fainter 
magnitudes, [O~III] $\lambda 5007$ is usually twice the strength of 
H$\beta$ \citep{p11, kim2}.  In the metal-poor environments of 
globular clusters, this lower limit is even more appropriate:  of the
11 halo and globular cluster PNe observed by \citet{howard97} and
\citet{jacoby+97}, all have an excitation parameter, 
$R = I(\lambda 5007)/I({\rm H}\beta) > 2$\null.  
In contrast, the vast majority of M31's H~II regions and 
diffuse ionized emission have H$\beta$ brighter than [O~III] $\lambda 5007$
\citep[\eg][]{blair, galarza, greenawalt}.  Moreover, in those
rare cases where an H~II region does exhibit a 
high value of $R$, its ionizing source (either a single O star or a very
young OB association) must be very hot.   Such an object will therefore
be very luminous --- more than 100 or 1000 times the brightness of a PN
central star --- and detectable either via its blue continuum
or its overly bright emission-line luminosity.

The only other sources that may have an excitation similar to that of a PN
are supernova remnants (SNRs), supersoft x-ray sources, and symbiotic stars.
Supernova remants are relatively rare (a factor of $\sim 10$ less numerous 
than PNe), and those remnants with $R > 2$ are less common still
\citep{magnier, galarza}.  If a SNR were embedded within one of our target 
clusters, it would likely be much brighter and/or have much broader emission 
lines than any PN\null. Supersoft x-ray sources are even rarer than SNRs, and 
most either have sizes much larger than a star cluster \citep{remillard} or
are associated with classical novae \citep{pietsch}.  In this latter
case, the nova ejecta would have a velocity structure that
is easily resolvable in our $R \sim 5000$~spectra.  Finally, symbiotic
stars can produce high-excitation emission lines, and in some cases, 
their observed properties can be very similar to those of true PN 
\citep{frankowski}.  In fact, \citet{symbiotic} has argued that {\it all\/} the
bright PNe seen in extragalactic surveys are actually symbiotic stars.  
But this is hard to prove, and PN surveys in the Milky Way and the LMC
find that symbiotic systems are only a minor contaminant \citep{iphas,
miszalski+11}.  Thus, emission-line sources that have 
[O~III] $\lambda 5007$ more than twice the strength of H$\beta$ are 
much more likely to be PNe than H~II regions, 
SNRs, or some other line-emitting object.

Another way of testing for unrelated line emission is through the
use of direct images.  Deep H$\alpha$ and [O~III] $\lambda 5007$ Mosaic CCD
frames of a 2.2~deg$^2$ region along M31's disk are available through
the Local Group survey program of \citet{massey}.  These images,
which reach point-source flux limits of $\log F \sim -15.7$ in 
H$\alpha$ and $\log F \sim -15.5$ in [O~III], can be used to examine
the immediate environment of each cluster.  Although not useful for
weak emission, the frames provided a check for
objects where the evidence for an associated PN was ambiguous.

\subsection{Associating a PN Candidate with a Star Cluster}

Even when high-excitation emission was detected in our fibers, its
source was not always associated with the underlying globular cluster. 
Some of the emission within M31's disk does have relatively high
excitation, and there is always the possibility of a chance
superposition of a cluster with a true but unassociated PN\null.  
We can quantify the latter likelihood by computing the probability that 
a field planetary nebula would be projected within $1\farcs 5$ of an
M31 cluster entirely by chance.  As in other galaxies, the distribution of 
M31 PNe closely follows that of the galactic light \citep{merrett}, and, from
the surface photometry of \citet{kent} and the bolometric corrections of
\citet{buzzoni}, we calculate that $\sim 2 \times 10^7 \, L_{\odot}$ of M31's 
diffuse luminosity (i.e., exclusive of the globular clusters) is projected 
within our survey fibers.  This 
number, coupled with M31's luminosity-specific PN density \citep{merrett}, 
and the planetary nebula luminosity function \citep{p2} implies the 
existence of $\sim 0.4$ superpositions in the top 2.5~mag of the 
PNLF, and $\sim 3$ unassociated PNe within the limits of our survey. 
Other high-excitation objects, such as old SNRs or supersoft
X-ray sources will then increase this number.  Clearly, we
cannot ignore the possibility that two rather rare objects can be 
projected within a single optical fiber.

The best way to reject these chance superpositions is to compare 
the radial velocity of each cluster's candidate PN to that of its stars.
In general, for a planetary nebula to be bound to a cluster, the
difference, $\Delta v$, between its emission-line radial velocity and the
absorption-line radial velocity of the cluster's stars should satisfy the 
criterion $\Delta v \lesssim 3 \sigma_{\rm eff}$, where $\sigma_{\rm eff}$ is 
the quadrature sum of the system's internal velocity dispersion and the 
uncertainty in the radial velocity measurements.  The former quantity exists 
for $\sim 60\%$ of the globular clusters in our survey, mostly through the
$R \sim 34,000$ MMT echelle spectroscopy of \citet{strader11}.  
For the remaining old stellar 
systems, we can estimate the line-of-sight velocity dispersions through the 
clusters' fundamental plane relation \citep[\eg][]{djorgovski, strader09, 
strader11}.  While we generally do not have access to information about a 
cluster's size or surface brightness, a projection of the \citet{strader11} 
clusters onto the $M-\sigma$ plane yields
\begin{equation}
\sigma({\rm km~s}^{-1}) \sim 23.5 + 8 M_{T_1} + 0.7 M_{T_1}^2 
\end{equation}
where $M_{T_1}$ is the cluster's absolute $T_1$ magnitude in the Washington 
system \citep{kim_lee}, and $\sigma$ is the observed velocity dispersion.
Typically, these velocity dispersions span the range $3 \lesssim 
\sigma_0 \lesssim 30$~\kms, with a median value near $\sim 8$~\kms.
Young clusters do not necessarily follow this relation, but from the
structural analysis by \citet{barmby+09}, their line-of-sight velocity
dispersion should be small, $\sigma_0 \lesssim 3$~\kms.

The second term which enters into $\sigma_{\rm eff}$ is that arising from 
the uncertainty of our velocity measurements.  This error has two
components.  The first, which is associated with our centroiding of the 
[O~III] $\lambda 5007$ emission line, is generally small: our velocity
measurements typically have errors that are less than $\sim 5$~\kms.  This is
in agreement with the results of \citet{kim2}, who used the same
telescope and instrument setup to obtain $\lesssim 5$~\kms\ precision
for faint PNe in distant galaxies.  

\begin{figure}
\figurenum{5}
\epsscale{1.0}
\plottwo{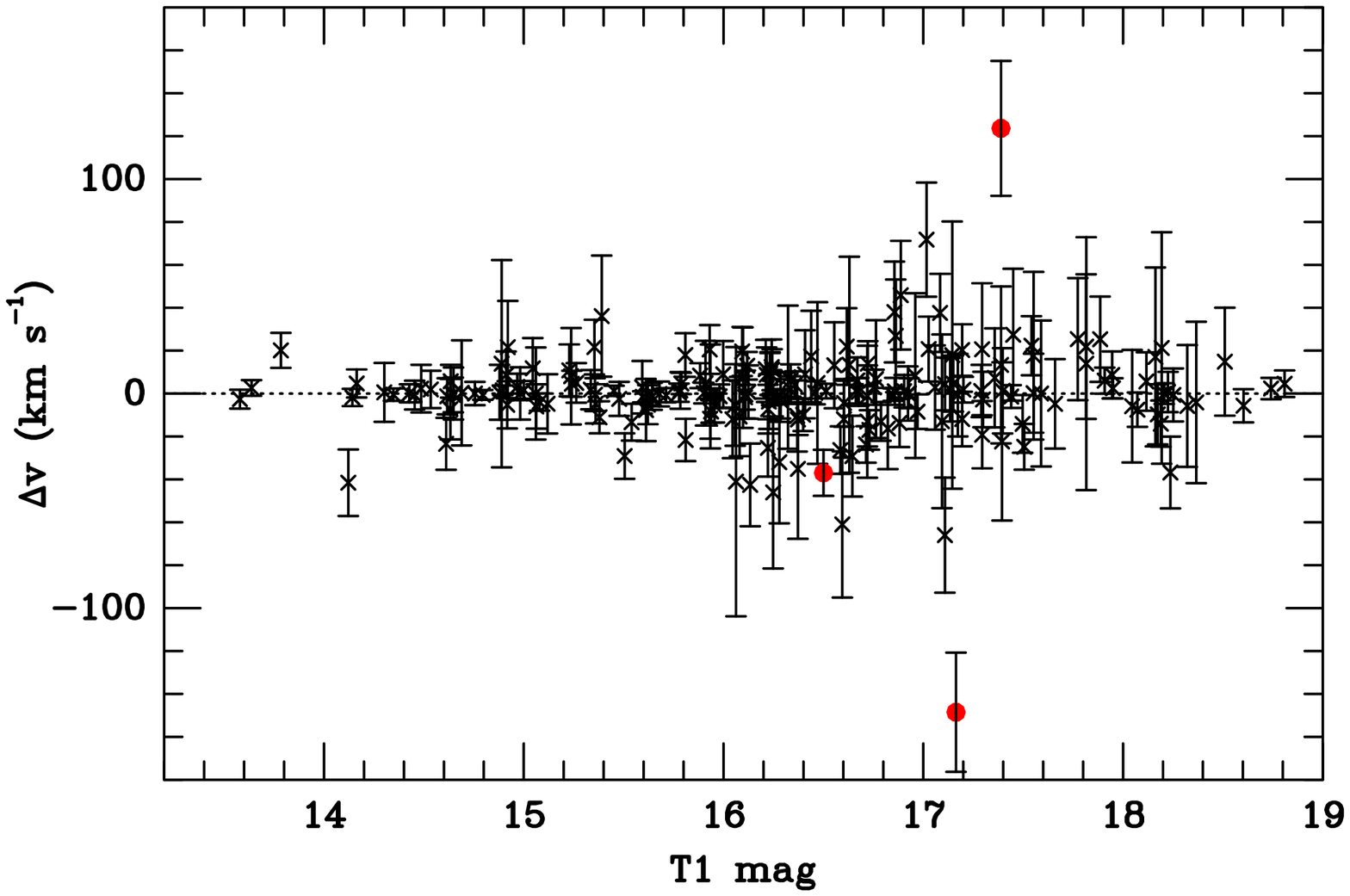}{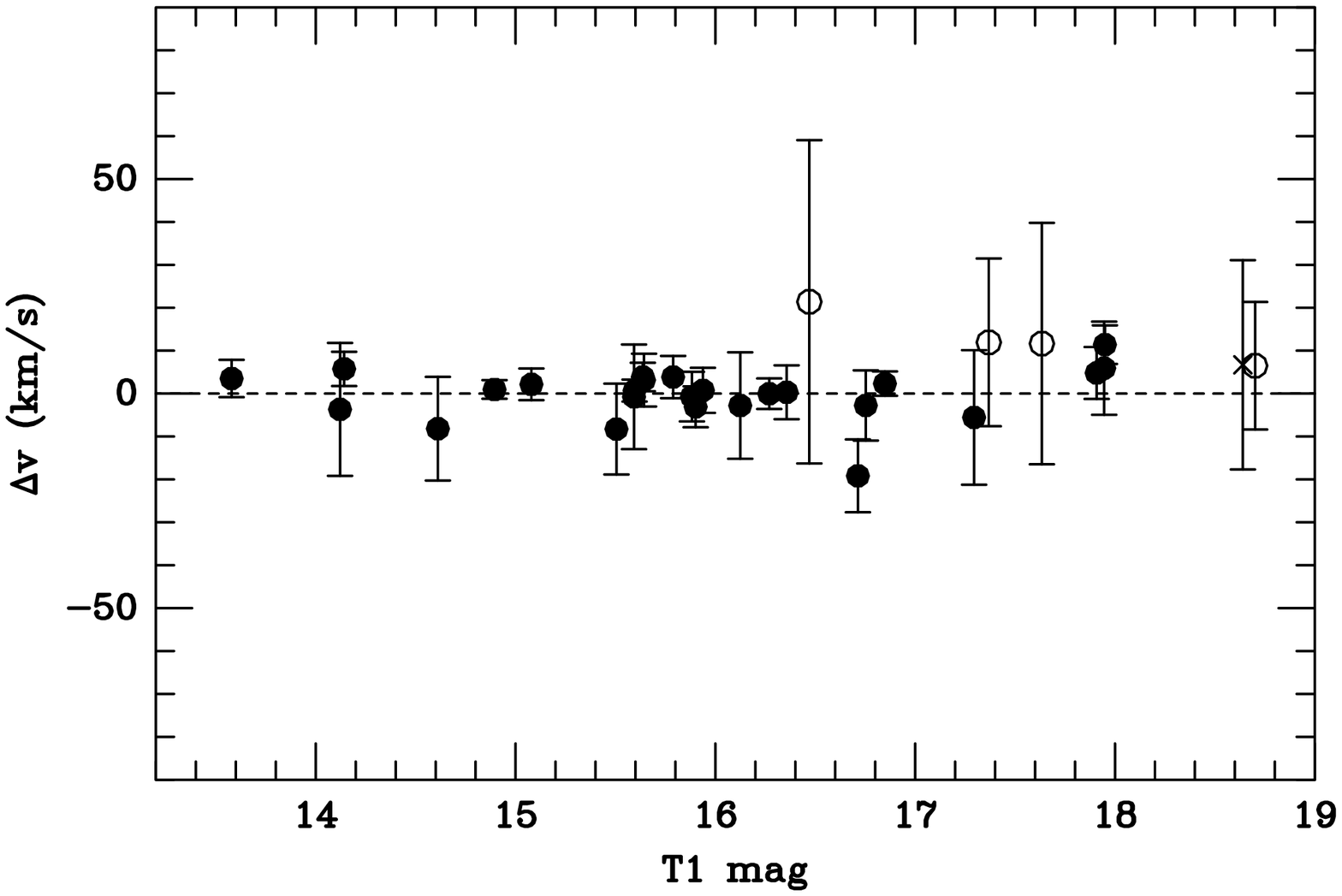}
\caption[rad_comparison]{Two estimates of the precision of our globular 
cluster velocities.  The left panel compares our data to those of the 
high-precision measurements of \citet{strader11}.  The solid red points denote
objects that differ from zero by more than three times the internal precision 
of the measurements; three additional systems have velocity discrepancies that
fall well off the plot.  The right panel shows the velocity
differences for clusters observed in more than one Hydra setup.
The scatter in these data is $7.6$~\kms. Solid points display old (globular) clusters;
open circles show younger systems.}
\label{rad_comparison}
\end{figure}

The other component of the error term, that coming from the absorption line
measurements, is more complex.  Almost half of our clusters have 
high-quality ($\sigma_v \lesssim 3$~\kms) velocity measurements, mostly 
through the MMT + Hectoechelle observations of \citet{strader11}.  As the
left panel of Figure~\ref{rad_comparison} shows, there 
is no systematic difference between our measurements and those of the 
Hectoechelle.  Six clusters have highly discrepant velocities, as they
differ from their \citet{strader11} values by more than 3 times the internal 
errors of the measurements.  Yet when these objects are removed, the
remaining 199 objects have a mean WIYN+Hydra velocity that is just 
$\Delta V = 0.2$~\kms\ greater than that of the Hectoechelle.  Since the
two sets of observations are on the same system, we can adopt the higher
precision \citet{strader11} velocities in our analysis.

For most of the remaining clusters, we can use our own velocity measurements, 
along with their associated measurement errors.  As the right panel
of Figure~\ref{rad_comparison} illustrates, the dispersion between our
independent velocity estimates of clusters observed in more than one Hydra
setup is consistent with the expectations of internal measurement error.  
Moreover, it is possible to obtain a quantitative estimate of our uncertainties
by comparing our velocity measurements (and their errors)
to those of the \citet{strader11} Hectoechelle data 
using the $\chi^2$ statistic
\begin{equation}
\chi^2 = \sum_i {(v - v_S)_i^2 \over 
(\sigma^2 + \sigma_S^2)_i }
\end{equation}
where $v$, $\sigma$, $v_S$, and $\sigma_S$ represent our velocities and their
uncertainties, and those of Strader, respectively.
When the six discrepant systems are removed, the reduced
$\chi^2$ for the 199 degrees of freedom is 0.96.  This strongly suggests that 
the internal errors of our velocity determinations are accurate, and can be 
adopted as the true uncertainties of our measurements.

Finally, as Figure~\ref{rad_comparison} shows, the typical error of our 
absorption line velocities is between 15 and 20~\kms.  However, a 
small number of objects have velocity uncertainties that are significantly
greater than this.  In a few cases, more accurate velocities are 
available from the literature, and in those cases, we either adopted the 
previously measured values, or averaged our velocities with the published 
data.   The remaining objects, where our velocity uncertainties were greater
than 30~\kms, were excluded from the analysis.  Those clusters for which
our velocities disagree with previously measured values by more than
four times the internal errors are given in Table~\ref{gc_verr}. 

\section{The Clusters Hosting Emission}

After forming $\sigma_{\rm eff}$, we excluded all PN candidates with 
[O~III] $\lambda 5007$ velocities that differed from that of their
parent cluster by more than $3 \, \sigma_{\rm eff}$.  We note that this
criterion may not remove all the false detections, since young clusters
are expected to have kinematics similar to that of the underlying
disk.  However, for the older systems, this should be a very 
effective discriminant.  Since the rotation-corrected velocity dispersion of 
M31's globular cluster system is $\sim 130$~\kms\ \citep{lee08}, the chance of 
finding a superposed disk source with the same velocity as one of our
candidate clusters is extremely low.

\begin{figure}
\figurenum{6}
\plotone{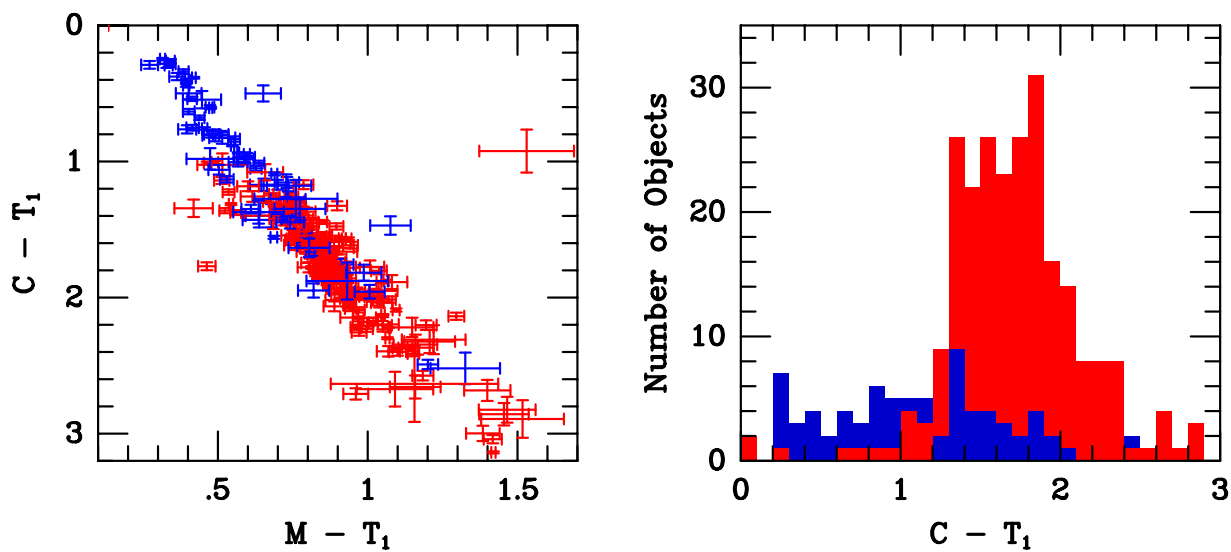}
\caption[2color]{Washington system photometry of M31 clusters classified by 
\citet{caldwell09, caldwell11}, with blue representing clusters with
ages $t \lesssim 2$~Gyr, and red designating globular clusters,
$t \sim 14$~Gyr.  On the left is a two-color diagram; on the right
is a histogram of $C - T_1$ colors.  In the absence of a spectroscopic
or {\sl HST\/} imaging age designation, we classify systems with
$C - T_1 < 1.2$ as young, and $C - T_1 > 1.2$ as old.}
\label{2color}
\end{figure}

Table~\ref{gc_summary} lists all the cluster candidates surveyed in this
program, along with their Washington system photometry obtained with the
KPNO 0.9-m telescope \citep{kim_lee}, their radial velocities, the 
equivalent widths of their [O~III] $\lambda 5007$ emission line, their
[O~III] $\lambda 5007$ to H$\beta$ emission-line ratio, and an age/type
classification.  For most of the clusters, this classification comes 
from the analyses of either \citet{caldwell09, caldwell11} or \citet{peacock}, 
and are based on a variety of measurements, including multi-bandpass 
photometry, {\sl HST\/} color-magnitude diagrams, and, in many cases, high 
signal-to-noise spectra.  To infer the ages of the 12 remaining objects
without any age classification, we used the clusters with known ages 
as a training set for our photometry. As Figure~\ref{2color} illustrates, 
most of the clusters designated as ``young'' are quite blue, with
$C - T_1 < 1.2$ and $M - T_1 < 0.7$.  Conversely, the overwhelming majority
of redder clusters are old.  Thus for the clusters without a previous
age determination, we can use color as a proxy for evolutionary status.
We classify any object redder than $C - T_1 = 1.2$ as old; bluer clusters
are identified as young.  These color-based classifications are designated
in Table~\ref{gc_summary} with a parenthesis.

Table~\ref{gc_cand} lists those M31 clusters which host possible 
planetary nebulae, along with the properties of the systems.  The 
signal-to-noise ratios for the [O~III] and H$\beta$ emission-line measurements 
are also tabulated.  The first three objects give the candidates associated 
with stellar systems confirmed as being old; the next five list candidates in 
the younger clusters.  Finally, two of our PN candidates are associated with 
controversial objects, \ie\  cluster candidates which may, in fact, be 
foreground stars. 
 
To estimate the [O~III] $\lambda 5007$ magnitudes of the PNe, we used our 
knowledge of the 5000~\AA\ continuum brightness of each cluster,
as determined by its Washington system $M$ magnitude (central 
wavelength 5075~\AA).  By measuring the strength of [O~III] $\lambda 5007$ 
relative to this continuum, \ie\ the line's equivalent width, we could
approximate the PN candidate's 5007~\AA\ monochromatic flux.  This flux
was then converted to a magnitude via
\begin{equation}
m_{5007} = -13.74 - 2.5 \log F_{5007}
\end{equation}
and placed on an absolute scale by assuming a distance of 750~kpc
\citep{keyfinal} and a differential extinction of $E(B-V) = 0.062$ 
\citep{schlegel}. On this scale, the brightest PNe in M31 attain a luminosity
corresponding to $M_{5007} = -4.5$ \citep{p2, merrett}. This value is
fairly resilient against alternative distant estimates to M31
\citep[e.g., 780~kpc;][]{mcc}.  

Of course, estimating brightnesses in this way carries a substantial amount
of uncertainty.  While the vast majority of M31 globular clusters have 
half-light radii smaller than the $1\farcs 5$ radius of our fibers, the tidal 
radii for these objects extend much farther \citep{barmby+07}.  As a result, 
the light coming through the fiber may not accurately reflect the total 
magnitude of the cluster, and astrometric errors in the cluster coordinates
and fiber positioning only exacerbate the problem.  

Moreover, the position of the emission-line source within the globular cluster 
is unknown.  Even if the fiber is centered on the globular cluster, a 
point-source planetary nebula may be offset by more than an arcsec.  For
example, if the four PNe within the Milky Way clusters were placed at the 
distance of M31, their typical separation from their cluster's center would be 
0.3-0.4\arcsec, but JaFu~1 in Pal~6 would be offset by a full 1.8\arcsec.  
This means that for three out of the four objects, the photometric error due 
to their position of the PN within the cluster would be negligible ($<1$\% 
loss through our 3\arcsec\  fibers), but in 1\arcsec\ seeing, JaFu~1's 
luminosity would be underestimated by a factor of $\sim 5$.  In the absence of 
high resolution imaging, our spectroscopic [O~III] luminosities are the best 
that can be achieved for these objects, and likely represent lower
limits to the true [O~III] $\lambda 5007$ brightnesses.  Other properties
which scale with luminosity, such as the inferred minimum central star
mass, would be lower limits as well.

\begin{figure}
\figurenum{7}
\plotone{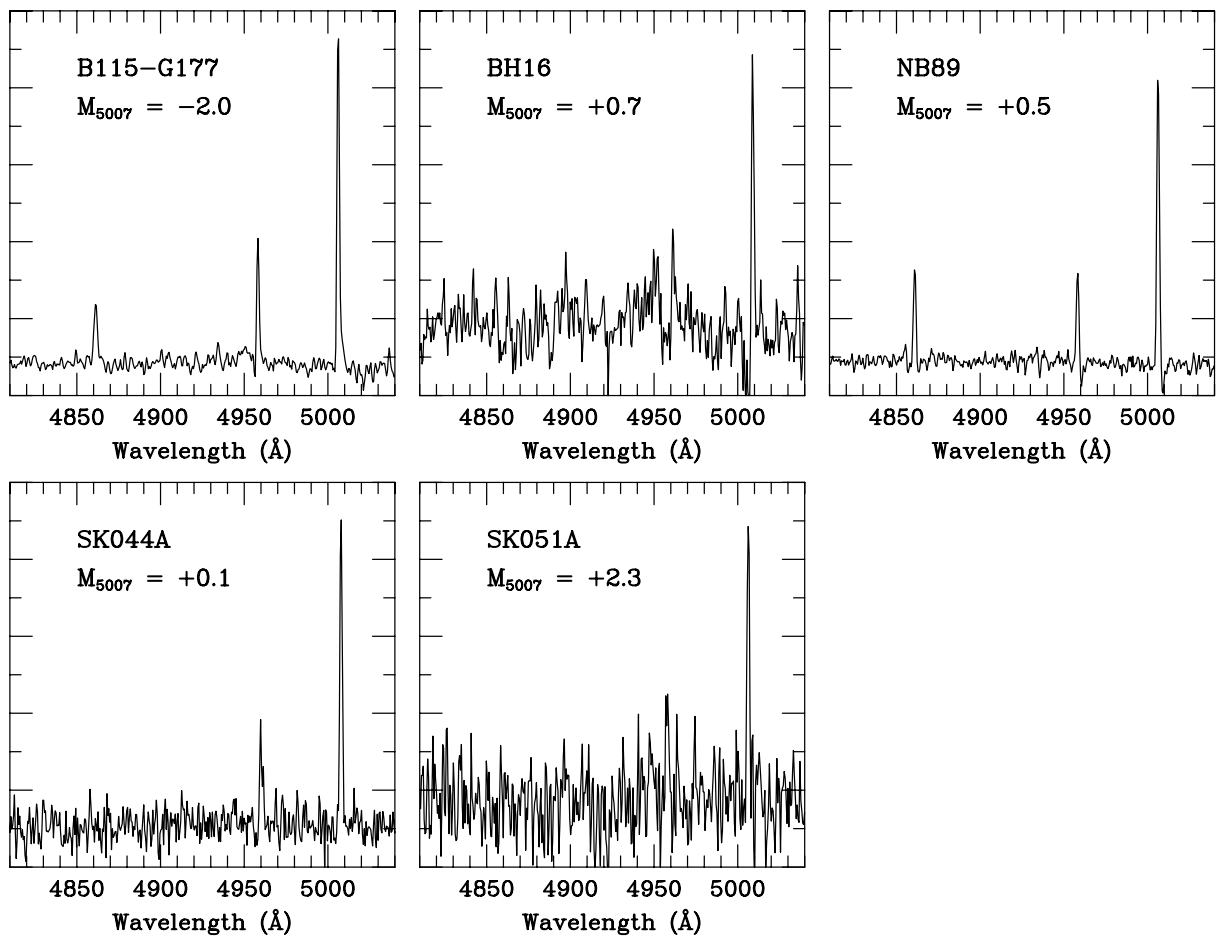}
\caption[gc_spectra]{The spectra of 5 systems with candidate planetary
nebulae in the wavelength range between 4800~\AA\ and 5100~\AA\null.
The top three panels show confirmed globular clusters; the bottom two
panels show objects which may be globulars.  To enhance the visibility of the 
emission lines, the spectrum of a template globular cluster has been subtracted 
from each object.  The y-axis represents counts; to convert to relative flux, 
note that the system response at H$\beta$ is roughly a factor of 1.46 less 
than that at [O~III] $\lambda 5007$.  For each candidate, the [O~III] line 
is at least twice the strength of H$\beta$.}
\label{gc_spectra}
\end{figure}

\begin{figure}
\figurenum{8}
\plotone{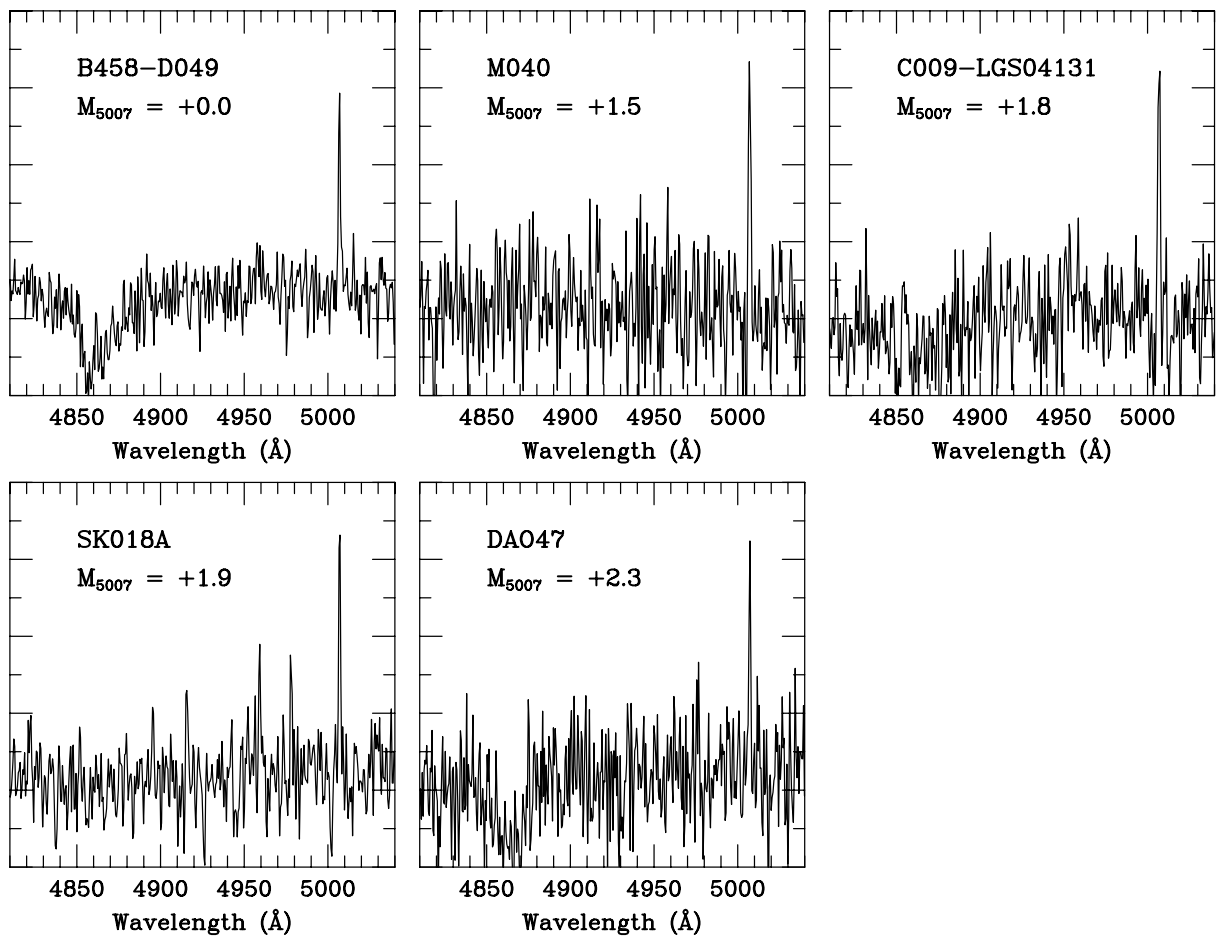}
\caption[yc_spectra]{The spectra of 5 M31 young clusters in the wavelength 
range between 4800~\AA\ and 5100~\AA\null.  To enhance the visibility of the 
emission lines, the spectrum of a template young cluster been subtracted from 
each object; in some cases, a mismatch in age has resulted in a poor
subtraction around H$\beta$.  The y-axis represents counts; to convert to 
relative flux, note that the response at H$\beta$ is roughly 1.46 times less 
than that at [O~III] $\lambda 5007$.  For each candidate, the [O~III] line 
is at least twice the strength of H$\beta$.}
\label{yc_spectra}
\end{figure}

The spectra of our candidate PN-GC associations are shown in 
Figure~\ref{gc_spectra} and Figure~\ref{yc_spectra}.
Below we detail their properties.

\subsection{Globular Clusters}

\citet{jacoby+97} argued that the single stars of old globular clusters cannot
form PNe due to the time scale of their post-AGB evolution:  by the time their
cores becomes hot enough for ionization, their ejected gas would have 
dispersed far into the interstellar medium.  Thus, \citet{jacoby+97} concluded
that PN formation inside globular clusters must involve binary stars,
either through mass transfer, which increases the core mass to that
of a higher mass progenitor, or through a binary interaction which 
accelerates the speed of post-AGB evolution \citep{moe12}.  More recently,
\citet{buell} has suggested an alternative, wherein the single stars of
globular clusters evolve high mass cores by being enriched in the helium
produced by previous generations of star formation in the cluster.  In any scenario,
however, central star mass is a critical parameter of the PN system, but 
one that is very difficult to determine, even for Galactic PNe.

We can, however, place limits on the mass of a PN central star using
models of post-AGB evolution.  At best, the [O~III] $\lambda 5007$
emission of a planetary nebula represents $\sim 10\%$ of the luminosity
being emitted from its central star \citep{dopita, schonberner10}.
Moreover, for any central star luminosity, there is a minimum core
required to generate that energy \citep{vw94, blocker}.  If we find that this 
minimum mass is too high to be produced by the evolution of a single star
of an old stellar population, then we will have strong evidence for a 
previous binary interaction.  We apply this approach to our candidate
objects.

{\it B115-G177:} According to \citet{caldwell11}, this globular cluster
is old and metal-rich, with [Fe/H] $\sim +0.1 \pm 0.1$.  Yet the system
harbors an emission-line source that is bright enough to stand out in
an Fe5015 versus Fe5270 plot, and hot enough
to have an [O~III] $\lambda 5007$ line that is three times the strength
of H$\beta$.  To place a lower limit on the luminosity of the exciting
source, we can use the fact that no more than $\sim 10\%$ of a central
star's total luminosity is reprocessed into [O~III] $\lambda 5007$
\citep{dopita, schonberner10}.  Consequently, for a source to be as bright
as $M_{5007} \sim -2.2$, its exciting star must be at least 
$\sim 700 \, L_{\odot}$ and have a mass of at least $\sim 0.54$~M$_{\odot}$.  
Although low-mass cores are capable of generating this amount of
luminosity, their evolutionary time scale is too slow to produce a 
planetary nebula.  This is our best candidate for a PN inside an M31 globular 
cluster and an object formed by binary evolution. It may be coincidental
that a rare PN is found in a relatively rare metal-rich, yet old, globular 
cluster \citep{woodley2010}, or perhaps the 
combination of properties is a clue to PN formation.

{\it BH16:}  This cluster, classified as old by \citet{strader11}, possesses
the X-ray source J004246.0+411736 \citep{fan}.  
Our velocity measurement for the
system is relatively poor ($-248 \pm 28$~\kms), due to possible
contamination from the underlying galactic bulge, and inconsistent with the
$-99.9 \pm 1.1$~\kms\ Hectoechelle measurement obtained by \citet{strader11}.  
If we adopt the latter value, then the velocity of the superposed [O~III] 
$\lambda 5007$ emission line differs from that of the cluster by less than 
20~\kms\ ($\sim 2.7 \sigma_{\rm eff}$), making an association likely.  The 
emission-line source itself has a high-excitation ($R \sim 4$), but
is also relatively faint, implying a central star luminosity that may be
as low as $100 \, L_{\odot}$.  

{\it NB89:}  The Lick indices \citep{gonzalez93} 
of this system imply an age of $\sim 10$~Gyr 
and a metallicity of [Z/H] $\sim -0.6$ \citep{barmby+00, beasley}, so the
object is most likely a globular cluster.  Its PN candidate is well-measured,
but faint, with an estimated [O~III] $\lambda 5007$ absolute magnitude
that is $\sim 4$~mag down the luminosity function.  The cluster also
barely satisfies our selection criteria:  [O~III] $\lambda 5007$ is just
2.1 times the strength of H$\beta$, and its inclusion in our list is partly 
due to the large ($\sim 30$~\kms) uncertainty in our estimate of cluster 
velocity.  If, instead of using our own measurement, we adopt the velocity 
of $-332 \pm 6$~\kms\ observed by \citet{barmby+00}, then the emission-line's 
velocity of $-384 \pm 5$~\kms\ is no longer consistent with it being part of 
the cluster. Caldwell (priv.~comm.)\ reports that forbidden emission from
[O~II] and [S~II] are strong, further suggesting that the emission-lines
are interstellar in nature.


\subsection{Candidate Globular Clusters}

{\it SK044A:}  \citet{caldwell09} classify this object as an M31 cluster
(of indeterminate age) based on its spectrum and its profile on archival
{\sl HST\/} frames.  In contrast, \citet{peacock} call the object a star, 
citing their analysis of UKIRT and SDSS data.  We also have difficulty
classifying the object: although the cluster's neutral color places it 
$0.062 \pm 0.056$~mag blueward of our ``young'' versus ``old'' dividing 
line, its measured radial velocity differs from that of M31's underlying 
disk by $\sim 100$~\kms\ \citep{chemin}.  Further complicating the
interpretation is our relatively poor determination for the cluster 
velocity ($-518 \pm 39$~\kms): even when combined with the $-491 \pm 46$~\kms\ 
measurement of \citet{kim_lee}, the resultant $\pm 29$~\kms\ uncertainty still 
dominates the error budget.  Nevertheless, the system is interesting, since its 
emission line velocity is inconsistent with a warm disk origin, and, with an
[O~III] $\lambda 5007$ to H$\beta$ ratio of $\gtrsim 12$, it has 
the highest emission-line excitation in our sample.  The [O~III] $\lambda 5007$
line is faint, so that if the emission is powered by a PN, its central star
could be fainter than $\sim 150 \, L_{\odot}$.  Unfortunately, without better
velocity information, we cannot say for certain whether the observed
48~\kms\ difference between the cluster's emission lines and absorption 
features is indicative of an association or a chance superposition. 
Caldwell (priv.~comm.)\ notes that [S~II] is weak in his spectrum, providing 
further support for the idea that the emission line is produced in a PN 
rather than the warm interstellar medium.

{\it SK051A:} This is our faintest PN candidate, and another object for
which we have a relatively large (22~\kms) measurement error.  Nevertheless,
our derived cluster velocity is in excellent agreement with that found by
\citet{kim_lee}, and the source does possess high-excitation ($R > 5$) 
[O~III] emission at a velocity consistent with both estimates.  The
system has the colors of an old cluster, but without better data, we
cannot confirm its association with the emission line.  \citet{peacock}
classify the object as a foreground star based on its appearance on
images from UKIRT and SDSS, and Caldwell (priv.~comm.)\ notes that the
object is not resolved on {\sl HST\/} frames.  Thus, it is possible
that the observed continuum is from a point source contaminant,
rather than a compact cluster.

\subsection{Young Clusters}

{\it B458-D049:} The cluster just barely satisfies our criterion, as the
velocity of the [O~III] $\lambda 5007$ emission-line differs from that of
the cluster by 18~\kms, or $2.9 \, \sigma_{\rm eff}$.  It is a young system,
as evidenced by the poor results of our template subtraction about the 
H$\beta$ absorption line, and \citet{caldwell09} estimate its age 
at 0.5~Gyr.  If the [O~III] emission does come from a planetary, then
the physics of single-star stellar evolution implies that the PN
is a high core-mass object.  Specifically, the relationship between
age and turnoff mass \citep{ibenlaughlin}, coupled with the initial
mass-final mass relation \citep{kalirai} yields $M_{\rm core} 
\sim 0.66 M_{\odot}$.  If the emission does come from a cluster PN,
then the object either evolved from a single massive (young) star
and is now well-past its peak [O~III] $\lambda 5007$ 
brightness of $\sim8000 L_{\odot}$, or it was created through
a binary pathway and is likely the result of common envelope 
evolution.  One can sometimes distinguish between these two
possibilities in the Galaxy where morphology and abundance
anomalies provide some discrimination \citep{miszalski+13}.  At the distance
of M31, luminosity is the primary indicator; if an object is very bright,
then it likely derives from a single massive star.

{\it M040:} \citet{caldwell11} re-classified this object as
young, though they did not estimate an age.  [O~III] $\lambda 5007$
is well-measured, [O~III] $\lambda 4959$ is weak, and $H\beta$ is
virtually undetectable.  The velocity agreement between the set of 
emission lines and that of the underlying cluster continuum is not
particularly good, between 13 and 28~\kms, depending on whether we
adopt our velocity or that of \citet{caldwell11}.  Given the
$\sim 30$~\kms\ uncertainties of both measurements, an association remains
a possibility.

{\it C009-LGS04131:}  Our velocity for this faint system is poor
($-483 \pm 51$~\kms) but it is in excellent agreement with the value of
$-495 \pm  32$~\kms\ measured by \citet{caldwell09} and with the velocity
of its [O~III] $\lambda 5007$ emission line ($-484 \pm 5$~\kms).  The cluster 
itself is young, with an age of $\sim 0.3$~Gyr \citep{caldwell09} and 
a turnoff mass of $\sim 3.1 M_{\odot}$ \citep{ibenlaughlin}.  Although [O~III] 
$\lambda 5007$ is rather faint, ($M_{5007} \sim +1.7)$, it is well-measured, 
and whatever is causing the emission has a very high excitation, $R > 6$.  If 
the exciting source is a planetary, then, based on the turnoff mass, it is 
either a $\sim 0.73 M_{\odot}$ core mass object which has faded substantially 
since its peak luminosity, or an object that was formed through a 
binary interaction. 


{\it SK018A:} We observed this young cluster twice, with 
consistent results ($\Delta v = 6.5$~\kms).  
\citet{caldwell09} estimate the age of the cluster to 
be $\sim 0.8$~Gyr, which implies a turnoff mass of $\sim 2.1 \, M_{\odot}$
\citep{ibenlaughlin} and a PN core mass of $\sim 0.62 \, M_{\odot}$
\citep{kalirai}.  Like C009-LGS04131, the object is more than
$\sim 6$~mag down the [O~III] $\lambda 5007$ PNLF. There is no evidence of
H$\beta$, which implies $R > 4$.

{\it DAO47:}  Our velocity, when combined with two other determinations in the
literature \citep{perrett02, caldwell09}, yields a value 
that is within 9~\kms\ of that measured for the [O~III] line.  The spectrum 
is relatively noisy, and even after template-subtraction, residual stellar 
H$\beta$ absorption is still visible.  Nevertheless, [O~III] $\lambda 5007$ is 
reasonably well-detected, and the narrow H$\beta$ emission line is completely 
absent.  \citet{caldwell09} estimate the age of the system to be 
$\sim 0.5$~Gyr, which, in terms of turnoff mass, is $\sim 2.5 M_{\odot}$ 
\citep{ibenlaughlin}.  The initial mass-final mass relation then implies 
$M_{\rm core} \sim 0.66 \, M_{\odot}$. 

\section{Discussion}

\subsection{Expected and Observed Numbers of PN Candidates}

In their survey of 130 Milky Way globular clusters, \citet{jacoby+97} 
identified four planetary nebulae.  Since our M31 survey targeted $\sim 270$
old clusters, a simple scaling of numbers suggests that we should have
found $\sim 8$~PNe in our survey.  However, only two of 
the Milky Way objects (Ps~1 in M15 and JaFu~1 in Pal~6) would have been 
definitively detected by our observations.  At the distance of M31, JaFu~2,
which is 7.3 mag down the PNLF, would be at, or just below, the threshold for
detection, and GJJC-1 in M22 (which may not be a true PN) would be well 
past our detection limit.  Consequently, we might expect to see 
$\sim 4$ objects; our list of systems with PN candidates contains 3 confirmed
globular clusters, and two other sources which may be old systems.  

It is unlikely that all of these candidates are true PNe, but the data are 
marginally consistent with the results of the Galactic surveys. Of the
5 candidates, 1 appears to be an excellent PN identification associated with 
an old cluster.  If the other 4 candidates are ultimately rejected, then the 
number of PNe in M31 globular clusters found in this study is 4 times lower 
than expected. This can be explained, in part, as a consequence of the 
observational challenges of the project.

Similarly, our data are consistent with the results of the \citet{peacock-4472} 
survey of the Virgo giant elliptical NGC~4472.  Their observations
of 174 luminous globular clusters covered $\sim 2 \times 10^8 L_{\odot}$
of bolometric light, and extended $\sim 2.5$~mag down the PN luminosity
function.  Our observations surveyed $\sim 2.5$ times less light, but extended
$\sim 3.5$~mag further down the PN luminosity function.  If we assume that 
the PNLF of globular clusters is similar to that of the field stars of
old populations, \ie
\begin{equation}
N(M) \propto e^{0.307 M} \{ 1 - e^{3 (M^* - M)} \}
\end{equation}
then the NGC~4472 observations imply that we should have seen
$\lesssim 2$~PN in our survey.  This, again, is consistent with our data.

Conversely, the number of PN candidates found in young clusters is
far more than anticipated.  The total luminosity of all the young clusters
included in our survey list is rather small, $M_V \sim -9.6$, and only
$M_V \sim -8.6$ was surveyed with good velocity precision. Because the
young clusters are members of the disk population, their velocities are
far more likely to match the velocities of
potential interlopers (e.g., HII regions, SNRs, disk emission) 
that are also in the disk, than would the old clusters. We therefore
must exercise additional caution when evaluating these candidate PNe.

If we assume a bolometric
correction of $-0.85$ \citep{buzzoni}, this implies that our survey of 
young clusters sampled $\sim 5 \times 10^5 L_{\odot}$ of light.  From the
theory of stellar energy generation, the bolometric luminosity stellar 
evolutionary flux for systems with ages between $\sim 0.1$ and $\sim 1$~Gyr
is $\sim 1 \times 10^{-11}$~stars~yr$^{-1}$~$L_{\odot}^{-1}$ \citep{renzini}.
Thus, we might expect these populations to produce one PN every 
$\sim 200,000$~years.  Even if these PNe remained visible for 50,000~yr, that 
is still not enough time to build up a detectable population.  It is therefore 
likely that the larger velocity errors associated with these fainter clusters, 
and the kinematic similarity of the clusters and the field exacerbate our
ability to discriminate PN candidates from superposed disk emission.

\subsection{Are Binary Stars Important}
\citet{straggler} have argued that blue stragglers are responsible
for most, if not all, of the bright PNe found in elliptical galaxies.
Not only do these objects possess the main-sequence masses needed to build 
high-mass post-AGB cores, but their evolutionary timescale, relative to that
of PN ($\sim 10^6$), is roughly the same as the relative numbers of
the two objects.  Since the creation of blue stragglers in globular
clusters may, in some way, be related to the rate of stellar encounters,
$\Gamma$ \citep[\ie][]{davies+04, leigh+11}, then it is at least possible 
that the probability of finding a PN would also be proportional to this 
factor.   From \citet{verbunt}, this means that 
\begin{equation}
N(PN) \propto \Gamma \propto {\rho_0^2 \, r_c^3 \over \sigma}
\label{eq:encounter}
\end{equation}
where $\rho_0$ is the cluster's central luminosity density, $r_c$ is
the core radius, and $\sigma$ is the stellar velocity dispersion.  The
structural parameters of Milky Way clusters \citep{harris96, harris10} 
do seem to support this idea, as two of the systems which host PNe 
have extremely high encounter rates (NGC~6441 has the second highest rate
of all Galactic GCs, and M15's rate ranks as seventh), and three
of the four rank in the top half of clusters.  Moreover, this result is not
simply due to the clusters having more stars: if one calculates the 
mass-specific encounter rates of the Milky Way clusters, then again,
three of the four clusters hosting PNe appear in the top half of the 
list.   The statistics of only four objects are poor, but the numbers
do suggest a connection between stellar encounters and PN formation.

Unfortunately, our emission-line survey of M31's globular cluster system does 
not yet allow us to increase the statistics significantly.  Although
over 250 of M31's globular clusters have structural measurements
\citep{barmby+07, peacock}, only one of the old clusters listed in 
Table~\ref{gc_cand} are included in that number.  Interestingly, if we assume 
that the clusters are in virial equilibrium (so that we can approximate the
encounter rate using $\Gamma \propto \rho_0^{1.5} r_c^2$), then the cluster in
question (B115-G177, which contains our best PN candidate) has a value of 
$\Gamma$ that ranks in the top $\sim 25\%$ of M31 systems.  This again is 
suggestive, but it cannot be considered definitive until the other systems 
listed in Table~\ref{gc_cand} are surveyed.

Alternatively, we can attempt to explore the frequency of stellar
encounters using X-ray emission as a proxy for encounter rate. 
\citet{pooley} have shown that in Galactic globular clusters, there is
an excellent correlation between $\Gamma$ and the number of low-mass X-ray 
binaries (LMXBs).  If each LMXB had the same luminosity, 
we could test the PN binary-formation hypothesis by searching for a 
correlation between PN presence and X-ray brightness.  \citet{jacoby+97}
did this experiment in the Milky Way, where each individual X-ray source can
be identified.  This led the authors to suggest that binaries were 
responsible for the cluster planetary nebulae.

In M31, one cannot resolve the individual X-ray sources
within each cluster, and counting the total X-ray emission from
a GC is not the same as measuring its total LMXB population.  
In fact, the large range of luminosities possible for LMXBs
makes any connection between X-ray luminosity and binary population
tenuous at best.  In our survey, $\sim 15\%$ of the globular clusters 
have X-ray sources, but only one appears in Table~\ref{gc_cand} 
\citep{stiele11}.  Consequently, X-ray emission in M31 clusters does 
not appear to present any evidence for the PN binary formation hypothesis.

\subsection{The Luminosities of the Candidates}
\begin{figure}
\figurenum{9}
\plotone{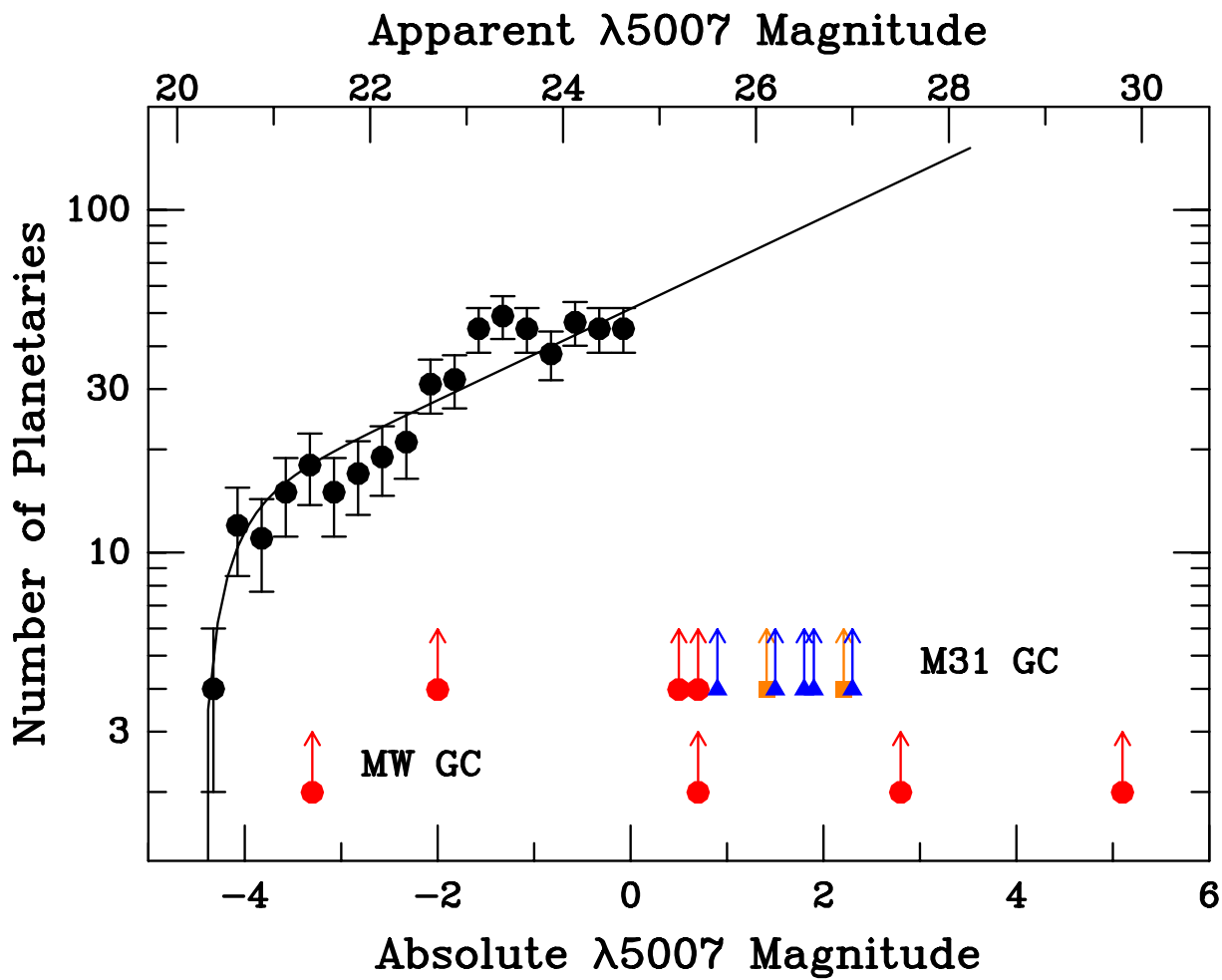}
\caption[m31_pnlf_gc]{The planetary nebula luminosity function for the bulge
of M31 (black points with error bars). These data, which extend over nearly
5~magnitudes, represent the deepest M31 PNLF currently available.  
The solid line shows the model PNLF (an exponential with a bright-end
cutoff) that was adopted by \citet{p2}.  The magnitudes of the clusters PNe
are marked: Milky Way PNe on the lower row, and M31 PNe on the upper row.
The red circles, blue triangles, and tan squares represent PN candidates
in old confirmed clusters, young clusters, and 
candidate globular clusters, respectively.}
\label{m31_pnlf_gc}
\end{figure}

Figure~\ref{m31_pnlf_gc} displays the [O~III] $\lambda 5007$ magnitudes
of planetary nebulae associated with Milky Way and M31 clusters.
From the figure, it is clear that most of the clusters found in our
spectroscopic survey are far fainter than those discovered via narrow-band
or counter-dispersed imaging.  This is simply a consequence of the
technique; slit spectroscopy reduces the sky background enormously, allowing
us to probe a region of M31's PN luminosity function that is undetectable by
other methods.

The more salient feature of the diagram is the distribution of relative
PN luminosities.  The systems of globular clusters in the Milky Way and M31 
each contain a lone bright PN candidate; the remaining objects are extremely
faint, in a regime where the luminosity function of LMC PNe is increasing
exponentially \citep{mash}.  This is expected if most faint PNe have
slowly evolving central stars embedded in freely expanding nebulae
\citep{hw63}.  In globular clusters, however, we might expect the PNLF to be
distorted, as the absence of intermediate mass stars might result in a deficit 
of intermediate luminosity PNe.  Thus far, however, there is no
evidence for this effect:  according to a Kolmogorov-Smirnov test, the combined
luminosity function of PNe in Milky Way and M31 globular clusters is
fully consistent with the curve displayed in Figure~\ref{m31_pnlf_gc}.
Of course, the numbers involved are small, but to date, there is no
reason to reject the hypothesis that the PNe of globular clusters obey
the simple law proposed by \citet{p2}.

\subsection{The Observational Challenge}

For Galactic clusters, searching for PNe is relatively straightforward.
As demonstrated by \citet{jacoby+97}, one can use the classic 
on-band/off-band technique to detect [O~III] $\lambda 5007$ over almost
the entire planetary nebula luminosity function.  Moreover, 
one can resolve both the cluster stars (except near the cluster center)
and, importantly, the nebula.  These two factors enable easy detection of
emission-line objects having the morphological characteristics of a PN, even
at very low surface brightnesses. Furthermore, relative to an extragalactic 
survey, cluster classifications are far more reliable, as are the velocity
measurements for both the PN and the cluster.  

In contrast, attempts to identify PNe in other galaxies are faced with a host
of complications.  Among these are:

{\it Mimics}: Objects other than PNe (\eg, H~II regions, SNRs, and diffuse
emission) can have similar spectral signatures over the limited wavelength
range of the WIYN Bench Spectrograph and its 740~lines~mm VPH grating.
This problem is partially technical, as many modern spectrographs
offer more complete spectral coverage at comparable resolution, allowing 
better discrimination against potential mimics.  Yet even in the Milky
Way, PN classifications can be controversial \citep[\eg][]{iphas, frew10}
and PN candidates are constantly being re-evaluated.  The limited
information available on extragalactic objects only exacerbates this
problem.

{\it Spectral resolution and sensitivity:} These instrumental parameters are 
probably the most critical factors for successfully and definitively 
finding PNe in extragalactic globular clusters.  As described above, our
ability to define the velocity (and the velocity dispersion) of the 
underlying star cluster severely limited our ability to exclude chance
superpositions.  We were fortunate that the literature provided an excellent
source of velocities for many of our objects.  For searches beyond M31, 
$\lesssim 3$~\kms\ velocity measurements will not always be available.
Similarly, sensitivity becomes increasingly important at larger distances, 
both for determining cluster velocities and for probing the faint end of
PNLF\null.  The study in NGC~4472 \citep{peacock-4472} only reached
2.5 mag down the PNLF; had we not gone far beyond this limit in M31, we 
likely would not have identified any PN candidates.  Finally,
resolution is also a helpful factor in this type of survey, both for
the detection of PN emission, and for eliminating mimics such as SNRs
and nova shells, which will have broad emission lines.  Again, this is a 
technical issue where large telescopes can dramatically advance studies like 
this one.

{\it Spectral aperture (slit width, fiber size):} 
Our fiber diameter was limited to 3\arcsec, which is a reasonable match
to the half-light radius of most M31 clusters.  Still, these fibers do not 
sample all of the cluster light, and outlying PN could be overlooked, either 
because they fall entirely outside the fiber, or near the fiber limits, where
flux is lost due to the effects of seeing.  One could, of course, choose
to use a large fiber or slit size for the observation (though at WIYN,
the largest fiber size is 3\arcsec), but this would reduce the 
signal-to-noise of the measurement by admitting more sky and galactic
background.  This problem is ameliorated when going to more distant galaxies, 
though again at the cost of a higher galactic background and a loss of 
sensitivity due to the greater distance.

{\it Multiple objects along the line of sight:} When observing a distant 
galaxy, there is a finite probability that two unrelated objects will fall
within the same spectroscopic aperture.  This problem becomes worse as the 
distance increases, as a given fixed aperture represents a broader spatial
swath of the galaxy.  To compensate for this effect, one needs better
velocity measurements so that unrelated objects can be discriminated.

In the future, searches for PNe in extragalactic globular clusters should be
more productive as many of these challenges can be overcome with technological
advancements. For example, the M31 problem becomes relatively easy with
the high signal-to-noise spectra produced by an extremely large (25-40~m
class) telescope (ELT) equipped with a multiobject, dual-channel, 
medium-resolution spectrograph.  Similarly, adaptive optics on ELTs
(or narrow-band filters on the {\sl Hubble Space Telescope\/} could be
used to resolve the cluster into stars, allowing the PN candidate to
be imaged directly.  This technique would not work as well for systems beyond
a few Mpc, but could produce a complete census of cluster PN in the
nearby universe.

\section{Conclusions}

We have demonstrated that it is possible to identify PN candidates in
distant globular clusters using spectroscopy around the [O~III] $\lambda 5007$
emission line.  The principal difficulty in this approach lies in confirming 
that the candidates are, indeed, associated with the cluster.  This requires 
precise radial velocity measurements, both for the emission line, 
and the underlying stellar continuum.  The better the velocity resolution of
the survey, the easier it is to separate embedded PNe
from chance superpositions, and to distinguish a cluster-bound PN from
other unrelated emission-line sources along the line of sight,
such as H~II regions, SNRs, and diffuse emission.

Of the 270 M31 globular cluster candidates observed with sufficient velocity 
precision, five show evidence for a candidate planetary nebula.  Given the
luminosity limits of the survey, the uncertainties in the velocity 
measurements, and the potential for confusion with other emission-line sources, 
this result is marginally consistent with the rate of PNe found 
in Milky Way clusters, \ie\ 3 or 4 in 130 clusters, and the upper
limit found in a survey of globular clusters around NGC~4472.  This rate 
argues that less than one-quarter of the stars in old clusters form PNe 
\citep{jacoby+97}. These numbers are also marginally
consistent with the binary hypothesis
for PN formation, which is about four times higher than the single star
production rate in old clusters \citep{moe}. 

At this time we cannot say how many cluster-associated PNe are likely to have
been formed from binary interactions.   Only one M31 PN candidate is embedded 
in a cluster whose structural parameters are known, but based on that cluster's
properties, and on the properties of the Milky Way's PN clusters, it appears
that the binary evolution scenario is a viable hypothesis to explain all 
globular cluster PNe.   High resolution imaging would help confirm
the existence of our PN candidates, and provide structural 
information on the host clusters.

In addition, we identified five PN candidates among the young clusters in 
our sample, a number that is much higher than that expected from the 
luminosity specific stellar death rate.  These are likely superpositions; most 
of the young systems surveyed have large velocity uncertainties, and the 
similarity between their kinematics and that of the underlying disk make it 
difficult to identify superposed objects. 

Finally, we emphasize that the M31 objects listed in Table~\ref{gc_cand}
are PN candidates only.  {\sl Hubble Space Telescope\/} narrow-band
images are needed to confirm their existence, especially for those
objects within young clusters.  However, once confirmed, these 
targets represent a new source of material for understanding the physics
of PN formation, and the chemistry of their parent clusters.

{\bf Acknowledgments}

We wish to thank Laura Fullton for early discussions about this project, and
Nelson Caldwell for helpful insights and providing access to his M31 cluster 
web page \break
http://www.cfa.harvard.edu/oir/eg/m31clusters/M31\_Hectospec.html). 
MGL is supported in part by the Mid-career Researcher Program through an 
NRF grant funded by the MEST (No.\ 2010-0013875). 
EK was supported by the NOAO/KPNO Research Experiences for Undergraduates 
(REU) Program, which is funded by the National Science Foundation (NSF) 
Research Experience for Undergraduates Program and the Department of Defense 
ASSURE program through Scientific Program Order No. 13 (AST-0754223) of the 
Cooperative Agreement No.\ AST-0132798 between the Association of Universities 
for Research in Astronomy (AURA) and the NSF.

\clearpage

\pagebreak
\clearpage

\pagebreak
\clearpage

\pagebreak
\clearpage

\pagebreak
\clearpage

\pagebreak
\clearpage

\pagebreak
\clearpage

\pagebreak
\clearpage

\pagebreak
\clearpage

\pagebreak
\clearpage

\begin{deluxetable}{lcccc}
\tablewidth{0pt}
\tablecaption{Log of Hydra Setups}
\tablehead{
&&\colhead{Exposure Times} 
&\multicolumn{2}{c}{Number of Targets} \\
\colhead{Setup} &\colhead{UT Date} &\colhead{(minutes)} 
&\colhead{GCs} &\colhead{Young Clusters} }
\startdata
M31-Y1 &2008 Oct 25 &$7 \times 30$ &71 &\phn 2  \\
M31-Y2 &2008 Oct 25 &$7 \times 30$ &68 &\phn 1  \\
M31-Y3 &2008 Oct 26 &$7 \times 30$ &57 &\phn 8  \\
M31-Y4 &2008 Oct 26 &$7 \times 30$ &43 &18      \\
M31-B1 &2008 Oct 27 &$7 \times 30$ &31 &25      \\
M31-G1 &2008 Oct 27 &$6 \times 30$ &37 &23      \\
M31-B2 &2008 Oct 28 &$7 \times 30$ &21 &26      \\
M31-R2 &2008 Oct 28 &$7 \times 30$ &16 &\phn 6  \\
\enddata
\label{obslog}
\end{deluxetable}
\clearpage

\begin{deluxetable}{lccccc}
\tablewidth{0pt}
\tablecaption{Globular Clusters with Discrepant Velocities}
\tablehead{
\colhead{Cluster} &\colhead{$T_1$} &\colhead{Our $v$ (\kms)} 
&\colhead{Published $v$ (\kms)} &\colhead{Source\tablenotemark{a}} 
&\colhead{Difference ($\sigma$)}  }
\startdata
B034D           &17.69  &$-37 \pm 29$    &$-347 \pm 25$      &1  &8.1  \\
B079D           &17.89  &$-89 \pm 29$    &$-394 \pm 25$      &1  &7.9  \\
B104-NB5        &16.29  &$-288 \pm 27$   &$32.9 \pm 0.7$     &2  &12.0 \\
B159            &18.67  &$-166 \pm 33$   &$-526.0 \pm 0.5$   &2  &10.8 \\
B451-D037       &17.75  &$-29 \pm 18$    &$-514 \pm 12$      &3  &22.3 \\
BH16            &17.16  &$-248 \pm 27$   &$-99.9 \pm 1.1$    &2  &5.3  \\
NB21-AU5        &17.22  &$-343 \pm 43$   &$-748.6 \pm 1.8$   &2  &9.4  \\
SK006A          &18.81  &$-25 \pm 38$    &$-320 \pm 54$      &4  &4.5  \\
SK021A          &18.25  &$-509 \pm 30$   &$-223 \pm 39$      &4  &5.8  \\
SK026A          &18.90  &$-62 \pm 33$    &$-568 \pm 55$      &4  &7.9  \\
SK045A          &15.50  &$+46 \pm 7$     &$-544 \pm 52$      &4  &11.2 \\
SK064A          &19.30  &$-35 \pm 16$    &$-259 \pm 42$      &4  &4.9  \\
SK079A          &18.02  &$-75 \pm 22$    &$-258 \pm 26$      &4  &5.3  \\
SK094A          &18.16  &$-158 \pm 21$   &$-8 \pm 29$        &4  &4.2  \\
SK096A          &17.94  &$-55 \pm 9$     &$-313 \pm 52$      &4  &4.9  \\
SK104A          &17.37  &$-171 \pm 10$   &$-301 \pm 17$      &4  &6.6  \\
SK106A          &18.94  &$-158 \pm 31$   &$-890 \pm 38$      &4  &14.9 \\
\enddata
\tablenotetext{a}{REFERENCES: (1) \citet{galleti06}; (2) \citet{strader11};
(3) \citet{perrett02}; (4) \citet{kim_lee} }
\label{gc_verr}
\end{deluxetable}
\clearpage

\begin{deluxetable}{lcccccccclcccl}
\tablewidth{0pt}
\tabletypesize{\scriptsize}
\rotate
\tablecaption{Globular Cluster Observations}
\tablehead{
&&&&&&&\colhead{Velocity} &\colhead{$\sigma_V$} & &  \\
\colhead{Name} &\colhead{$T_1$} &\colhead{$\sigma_{T_1}$\tablenotemark{a}}
&\colhead{$C$\tablenotemark{a}} &\colhead{$\sigma_C$} 
&\colhead{$M$\tablenotemark{a}} &\colhead{$\sigma_M$}
&\colhead{(\kms)} &\colhead{(\kms)} &\colhead{Type\tablenotemark{b}} 
&\colhead{Ref\tablenotemark{b}} &\colhead{$R$\tablenotemark{c}} 
&\colhead{EW (\AA)} &\colhead{Notes}
}
\startdata
AU010          &16.555 &0.017 &18.325 &0.031 &17.388 &0.019 &$-311.7$ &20.1 &Old   &1  & 2.8  &0.8    & \\
B001-G039      &16.395 &0.005 &18.580 &0.013 &17.377 &0.007 &$-202.7$ &10.9 &Old   &3  &\dots &\dots  & \\
B003-G045      &17.151 &0.012 &18.564 &0.014 &17.885 &0.010 &$-376.7$ &13.0 &Old   &1  &\dots &\dots  & \\
B004-G050      &16.452 &0.007 &18.172 &0.010 &17.255 &0.006 &$-378.1$ &5.9  &Old   &3  &\dots &\dots  & \\
B005-G052      &15.045 &0.002 &16.901 &0.004 &15.905 &0.002 &$-271.6$ &14.1 &Old   &1  &\dots &\dots  & \\
B006-G058      &14.967 &0.002 &16.780 &0.003 &15.831 &0.002 &$-232.9$ &6.1  &Old   &1  &\dots &\dots  & \\
B006D-D036     &17.634 &0.016 &18.386 &0.013 &18.013 &0.013 &$-538.1$ &28.1 &Young &3  &\dots &\dots
&Mean; $\Delta v = 11.6$~\kms \\
B008-G060      &16.270 &0.006 &18.077 &0.009 &17.103 &0.005 &$-324.6$ &3.6  &Old   &1  &\dots &\dots
&Mean; $\Delta v = 0.0$~\kms \\
\enddata
\tablenotetext{a}{Numbers in parenthesis have been transformed into
the Washington system using literature $BVI$ measurements and the 
relations of \citet{bessell}.}
\tablenotetext{b}{The cluster age/type classification has been adopted from the 
following references:  (1) \citet{strader11}; (2) \citet{caldwell11};
(3) \citet{caldwell09}; (4) \citet{peacock} }
\tablenotetext{c} {Derived ratio of [O~III] to H$\beta$ using the procedures
described in \S~2}
\tablenotetext{d}{Classified on the basis of an emission line a 5046~\AA.}
\tablenotetext{e}{Classified on the basis of a 50~\AA\ FWHM 
logarithmic-profile emission line at 4836~\AA.}
\tablenotetext{f}{Classified on the basis of weak emission at 4548~\AA.}
\tablenotetext{g}{Both [O~III] $\lambda 5007$ and H$\beta$ are well resolved
and have the same double-peaked emission profile.}
\tablenotetext{h}{Classified using the redshifted emission lines of H$\beta$
and the [O~III] doublet.}
\tablenotetext{i}{Classified as a galaxy in the literature, but our spectrum
is consistent with the object being a star cluster.}
\label{gc_summary}
\end{deluxetable}

\begin{deluxetable}{lccccccccc}
\tablewidth{0pt}
\rotate
\tablecaption{Clusters with Candidate Planetary Nebulae}
\tablehead{
\colhead{Cluster} &\colhead{Age (Gyr)\tablenotemark{a}} &\colhead{$M_V$} 
&\colhead{S/N (5007~\AA)} &\colhead{S/N (H$\beta$)} 
&\colhead{$R$\tablenotemark{b}} &\colhead{EW (\AA)} &\colhead{$M_{5007}$} 
&\colhead{$\Delta v$~(\kms)} &\colhead{$\Delta v / \sigma_{\rm eff}$} }
\startdata
\multicolumn{8}{l}{Globular Clusters} \\
B115-G177    &14      &$-8.54$  &72  &16  &3.0  &2.5   &$-2.0$ &$-5$  &0.42 \\
BH16         &Old     &$-6.46$  &8.3 &2.4 &4.0  &0.7   &$+0.7$ &$-19$ &2.71 \\
NB89\tablenotemark{c} 
             &10.4    &$-6.60$  &36  &12  &2.1  &1.5   &$+0.5$ &$-34$ &1.10 \\
\\
\multicolumn{8}{l}{Young Clusters} \\
B458-D049    &0.5     &$-6.73$  &6.0 &1.6 &2.1  &0.8   &$+0.9$ &$+18$ &2.87 \\
M040         &Young   &$-4.98$  &5.5 &1.1 &4.1  &1.6   &$+1.5$ &$+24$ &0.91 \\
\enddata
\label{gc_cand}
\tablenotetext{a}{Age estimate of \citet{caldwell09, caldwell11}; Objects
labeled as ``Star'' were classified by \citet{peacock}.}
\tablenotetext{b} {Ratio of [O~III] to H$\beta$}
\tablenotetext{c}{Age from \citet{beasley}; emission may be warm ISM}
\end{deluxetable}

\end{document}